\pdfoutput=1
\documentclass[cernpreprint,english]{na61doc}

\usepackage{lineno}
\usepackage{enumerate}
\usepackage{todonotes}

\usepackage{colortbl}
\definecolor{darkred}{rgb}{0.5,0,0}
\definecolor{darkblue}{rgb}{0,0,0.5}
\definecolor{firebrick}{rgb}{0.75,0.125,0.125}
\definecolor{darkgreen}{rgb}{0,0.5,0}

\usepackage[colorlinks=true,linkcolor=firebrick,citecolor=darkgreen,urlcolor=darkblue]{hyperref}

\newcommand{\eV}{\ensuremath{\mbox{e\kern-0.1em V}}\xspace}
\newcommand{\GeV}{\ensuremath{\mbox{Ge\kern-0.1em V}}\xspace}
\newcommand{\TeV}{\ensuremath{\mbox{Te\kern-0.1em V}}\xspace}
\newcommand{\MeV}{\ensuremath{\mbox{Me\kern-0.1em V}}\xspace}
\newcommand{\GeVc}{\ensuremath{\mbox{Ge\kern-0.1em V}\!/\!c}\xspace}
\newcommand{\AGeV}{\ensuremath{A\,\mbox{Ge\kern-0.1em V}}\xspace}
\newcommand{\AGeVc}{\ensuremath{A\,\mbox{Ge\kern-0.1em V}\!/\!c}\xspace}
\newcommand{\MeVc}{\ensuremath{\mbox{Me\kern-0.1em V}/c}\xspace}

\newcommand{\cm}{\ensuremath{\mbox{cm}}\xspace}

\newcommand{\dd}{\ensuremath{{\textrm d}}\xspace}
\newcommand{\dedx}{\ensuremath{\dd E\!/\!\dd x}\xspace}

\newcommand{\pt}{\ensuremath{p_{\textrm T}}\xspace}

\newcommand{\mt}{\ensuremath{m_{\textrm T}}\xspace}
\newcommand{\pp}{\mbox{\textit{p+p}}\xspace}

\newcommand{\y}{\ensuremath{{y}}\xspace}

\newcommand{\pim}{\ensuremath{\pi^-}\xspace}
\newcommand{\pip}{\ensuremath{\pi^+}\xspace}

\newcommand{\Xim}{\ensuremath{\Xi^-}\xspace}
\newcommand{\Xip}{\ensuremath{\overline{\Xi}^+}\xspace}
\newcommand{\Xir}{\ensuremath{{\Xi\left(1530\right)^{0}}}\xspace}
\newcommand{\Xirb}{\ensuremath{\overline{\Xi}\left(1530\right)^{0}}\xspace}

\newcommand{\Urqmd}{{\scshape U}r{\scshape qmd}\xspace}

\newcommand{\Epos}{{\scshape Epos}\xspace}

\newcommand{\CernVM}{\textsc{Cern\-\kern-0.05emVM}\xspace}

\ShineTitle{Measurements of \Xir and \Xirb production in proton-proton interactions at $\sqrt{s_{NN}}$~=~17.3~\GeV in the \NASixtyOne experiment}

\PreprintIdNumber{CERN-EP-2021-078}

\ShineJournal{EPJ C}
\ShineAbstract
{
Double-differential yields of $\Xir$ and $\Xirb$ resonances produced in \pp interactions were measured at a laboratory beam momentum of 158~\GeVc. This measurement is the first of its kind in \pp interactions below LHC energies. It was performed at the CERN SPS by the \NASixtyOne collaboration. Double-differential distributions in rapidity and transverse momentum were obtained from a sample of 26$\cdot$10$^6$ inelastic events. The spectra are extrapolated to full phase space resulting in mean multiplicity of $\Xir$ (6.73 $\pm$ 0.25 $\pm$ 0.67)$\times10^{-4}$ and $\Xirb$ (2.71 $\pm$ 0.18 $\pm$ 0.18)$\times10^{-4}$.
The rapidity and transverse momentum spectra and mean multiplicities were compared to predictions of string-hadronic and statistical model calculations.
}
\begin{document}
\maketitle
\section{Introduction}

Double-differential yields of $\Xir$ and $\Xirb$ resonances were measured in inelastic \pp interactions at laboratory beam momentum of 158~\GeVc. The measurement was performed at the CERN SPS by the \NASixtyOne collaboration~\cite{Abgrall:2014fa}.

 The description of strange quark production in hadron-hadron interactions and their subsequent hadronization are challenging tasks for QCD-inspired and string-based phenomenological models. This applies especially to doubly strange hyperons and their resonances. No experimental results are available from measurements of \pp interactions in the CERN SPS energy range. 
 At higher energy, $\Xir$ and $\Xirb$ resonances were studied in \pp interactions at the CERN LHC~\cite{Abelev:2014qqa}. Microscopic string/hadron transport models are widely used to describe and understand relativistic heavy-ion collisions. Data on strangeness production and especially on doubly strange resonances in \pp interactions provide important input data for these models.   

The paper is organised as follows. The \NASixtyOne detector system is presented in Sec.~\ref{sec:setup}. Sections~\ref{sec:method}-\ref{sec:systematics} are devoted to the description of the analysis method. The results are shown in Section~\ref{sec:results} and compared to published data and model calculations in Sec.~\ref{sec:comparison}. Section~\ref{sec:summary} closes the paper with a summary and outlook.

The following variables and definitions are used in this paper. The particle rapidity \textit{y} is calculated in the \pp center of mass system (cms), $y=0.5ln[(E+cp_L)/(E-cp_L)]$, where $E$ and $p_L$ are the particle energy and longitudinal momentum, respectively. The transverse component of the momentum is denoted as $p_T$. The momentum in the laboratory frame is denoted $p_{lab}$ and the collision energy per nucleon pair in the centre of mass by $\sqrt{s_{NN}}$.The unit system used in the paper assumes $c$=1.

\section{Setup of \NASixtyOne Experiment}\label{sec:setup}

\begin{figure}
\centering
\includegraphics[width=0.9\textwidth]{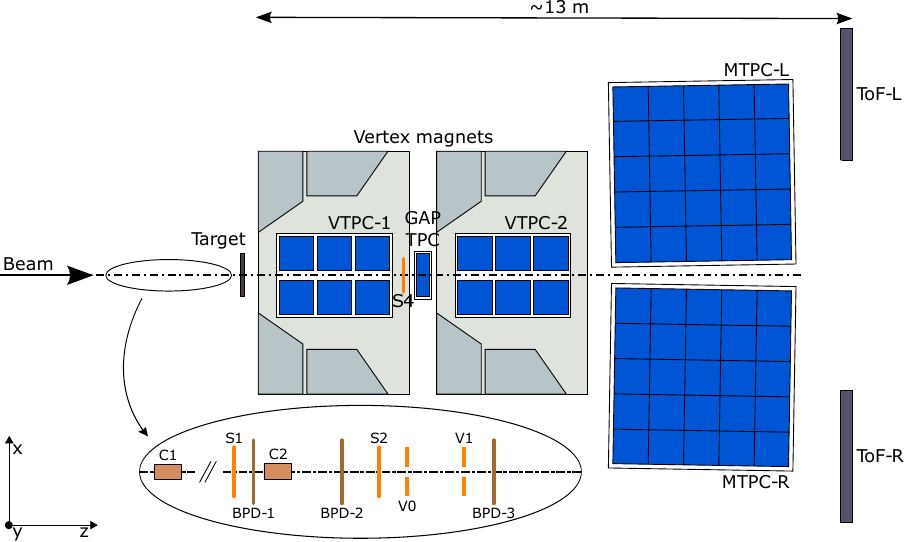}
\caption{(Color online) Schematic layout of the NA61/SHINE experiment
at the CERN SPS (horizontal cut, not to scale) showing the detectors used in the data taking.
The orientation of the \NASixtyOne coordinate system is shown in the picture. 
The nominal beam direction is along the z-axis. The magnetic field bends charged particle
trajectories in the x-z plane. The electron drift direction in the TPCs is along the y (vertical) axis. 
}
\label{fig:detector-setup}
\end{figure}

Data used for the analysis were recorded at the CERN SPS accelerator complex with the \NASixtyOne fixed target large acceptance hadron detector~\cite{Abgrall:2014fa}. The schematic layout of \NASixtyOne detector system is shown in Fig.~\ref{fig:detector-setup}. The \NASixtyOne tracking system consists of 4 large volume time projection chambers (TPCs). Two of the TPCs (VTPC1 and VTPC2) are within superconducting dipole magnets. Downstream of the magnets, two larger TPCs (MTPC-R and MTPC-L) provide acceptance at high momenta. The fifth small TPC (GAP-TPC) is placed between VTPC1 and VTPC2 directly on the beamline. The interactions were measured in the H2 beamline in the North Experimental Hall with a secondary beam of 158~\GeVc positively charged hadrons impinging on a cylindrical Liquid Hydrogen Target (LHT) of 20~cm length and 2~cm diameter. 
This beam was produced by 400~\GeVc protons hitting a Be-target. The primary protons were extracted from the SPS in a slow extraction mode with a flat-top lasting about 10 seconds. Protons and other positively charged particles produced in the Be-target constitute the secondary hadron beam. Two Cherenkov counters identified the protons, a CEDAR (either CEDAR-W or CEDAR-N) and a threshold counter (THC). A selection based on signals from the Cherenkov counters identified the protons with a purity of about 99\%~\cite{Abgrall:2013qoa}. The beam momentum and intensity was adjusted by proper settings of the H2 beamline magnet currents and collimators. A set of scintillation counters selects individual beam particles (see inset in~Fig.\ref{fig:detector-setup}). Their trajectories are precisely measured by three beam position detectors (BPD-1, BPD-2, BPD-3)~\cite{Abgrall:2014fa}. 

\section{Event selection}\label{sec:method}

A total of 53 million minimum bias \pp events were recorded in 2009, 2010, and 2011 and analysed. Interactions in the target are selected with the trigger system by requiring an incoming beam proton and no signal from the counter S4 placed on the beam trajectory between the two vertex magnets (see Fig.~\ref{fig:detector-setup}). 

Inelastic \pp events were selected using the following criteria:
\begin{enumerate}[(i)]
\item no off-time beam particle detected within a time window of $\pm$2$~\mu$s around the time of the trigger particle,
\item beam particle trajectory measured in at least three planes out of four of BPD-1 and BPD-2 and in both planes of BPD-3,
\item the primary interaction vertex fit converged,
\item z position of the interaction vertex (fitted using the beam trajectory and TPC tracks) not farther away than 9~\cm from the centre of the LHT,
\item events with a single, positively charged track with laboratory momentum close to the beam momentum (see Ref.~\cite{Abgrall:2013qoa}) were rejected, eliminating most of the elastic scattering reactions.
\end{enumerate}

After the above selection, 26 million inelastic events remain for further analysis.

\section{Reconstruction and selection of \Xir and $\Xirb$ candidates}

Reconstruction started with pattern recognition, momentum fitting, and the formation of global track candidates.  These track candidates generally span multiple TPCs and are generated by charged particles produced in the primary interaction and at secondary vertices.  

Particle identification was performed via measurement of the specific energy loss (\dedx) in the TPCs.
The achieved \dedx resolution is 3–6\% depending on the reconstructed track length~\cite{Abgrall:2014fa, Aduszkiewicz:2017sei}. The dependence of the measured \dedx on velocity was fitted to a Bethe-Bloch type parametrization.

\begin{figure}
\centering
\includegraphics[width=0.5\textwidth]{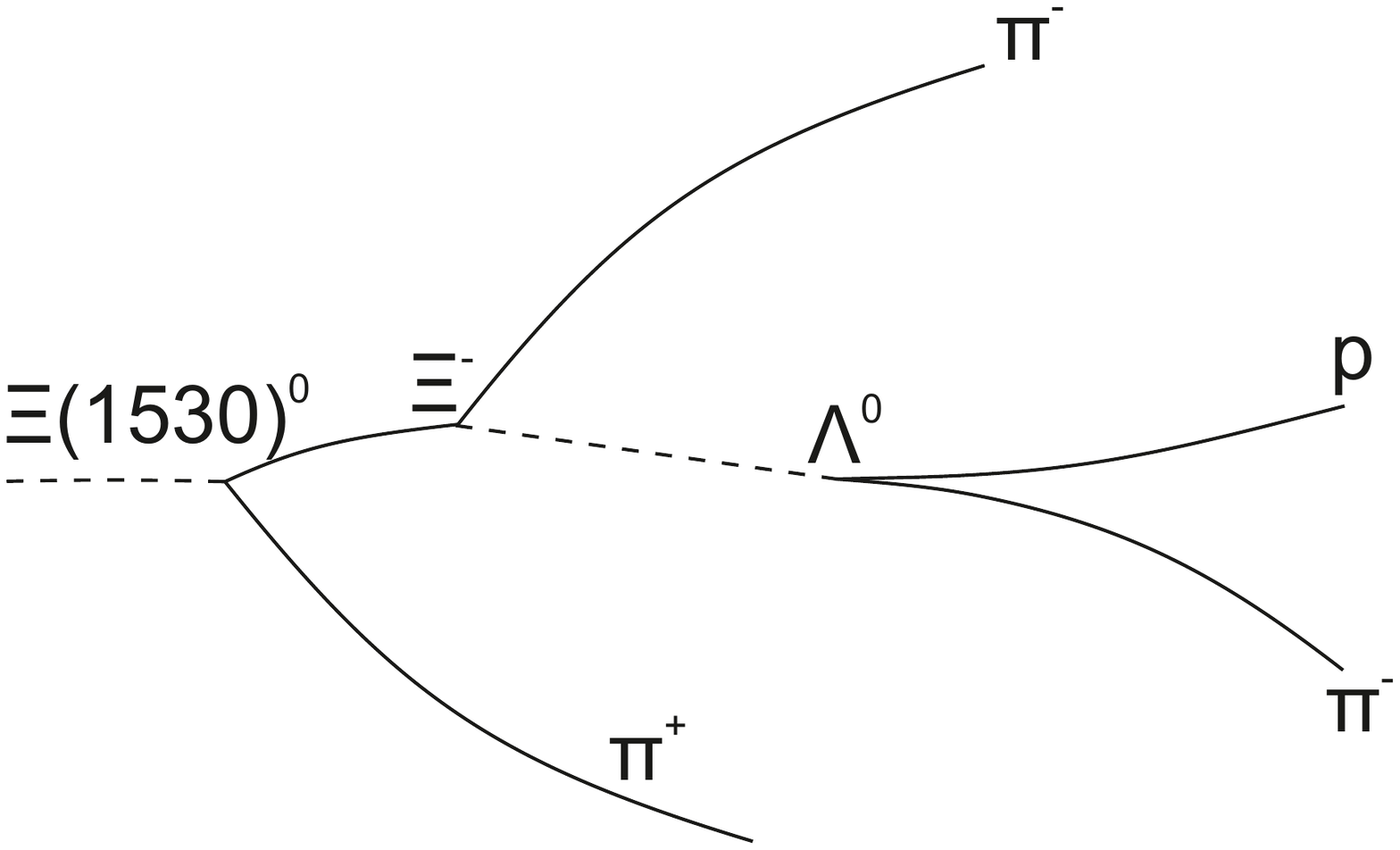}
\caption{
Schematic sketch of the \Xir decay scheme.
}
\label{fig:idea}
\end{figure}

The \Xir is produced in the primary interaction and decays strongly into \Xim and \pip. Then the \Xim travels for some distance, after which it decays into a $\Lambda$ and a \pim. Subsequently, $\Lambda$ decays into a proton and a \pim. A schematic drawing of the \Xir decay chain is shown in Fig.~\ref{fig:idea}.

The first step in the analysis was the search for $\Lambda$ candidates, which were then combined with a \pim to form the \Xim candidates. Next, the $\Xir$ was searched for in the \Xim\pip invariant mass spectrum, where the \pip originates from the primary vertex. An analogous procedure was followed for the antiparticles.

The $\Lambda$ candidates are formed by pairing reconstructed and identified tracks with appropriate mass assignments and opposite charges. These particles are tracked backwards through the \NASixtyOne magnetic field from the first recorded point, which is required to lie in one of the VTPC detectors. This backtracking is performed in 2 cm steps in the z (beam) direction. Their separation in the transverse coordinates x and y is evaluated at each step, and a minimum is searched for. A pair is considered a $\Lambda$ candidate if the minimum distance of the closest approach in the x and y directions is below 1 cm in both directions. Using  the track parameters and the distances at the two neighbouring space points around the point of the closest approach, a more accurate $\Lambda$ decay position is found by interpolation. This position, together with the momenta of the tracks at this point, is used as the input for a 9 parameter fit using the Levenberg-Marquardt fitting procedure~\cite{press_etal_1992}. 

\begin{figure*}[t!]
\centering
\includegraphics[width=.43\textwidth]{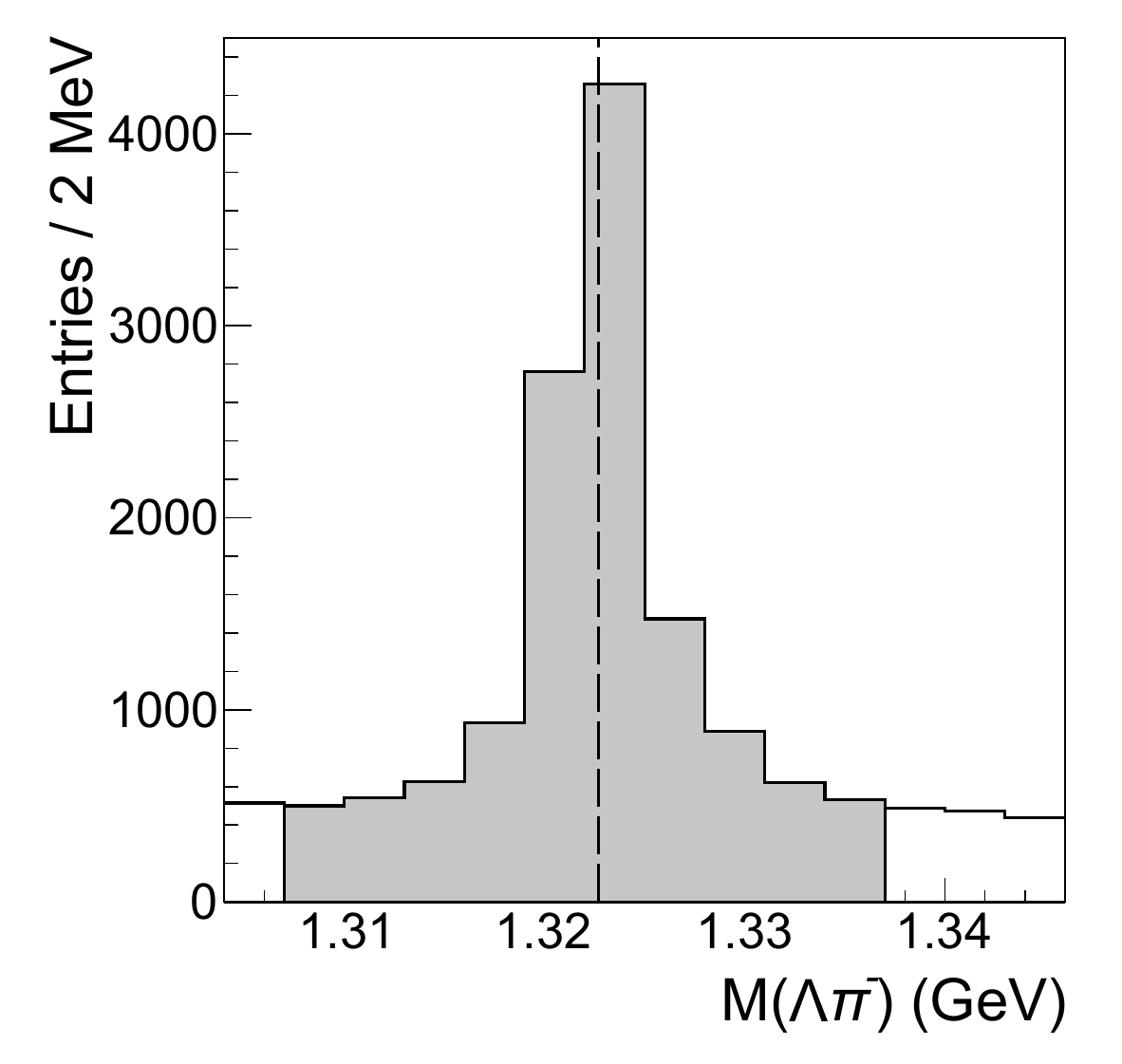}
\includegraphics[width=.43\textwidth]{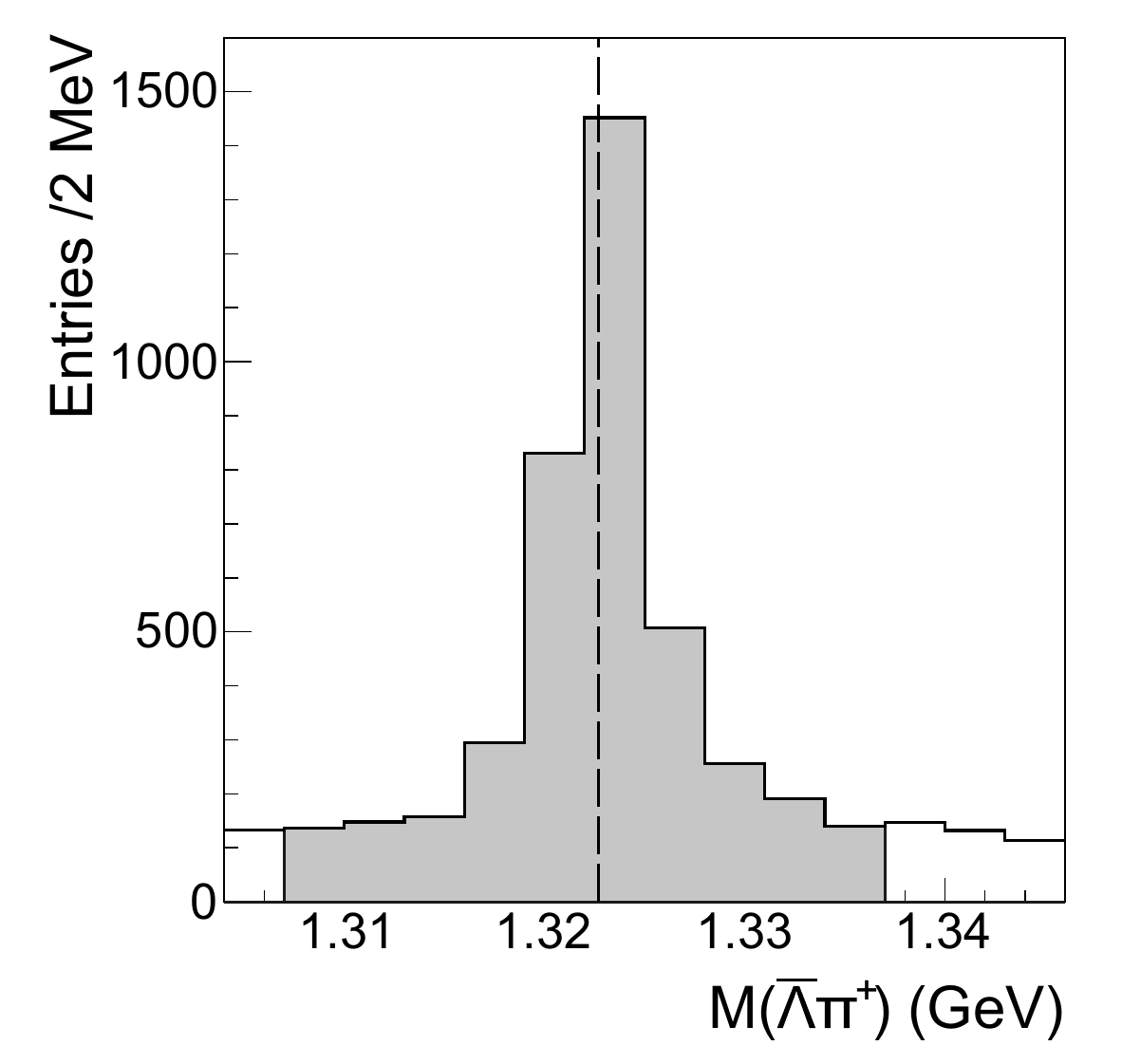}
\caption{
The $\Lambda$\pim ($\overline{\Lambda}$\pip) invariant mass spectrum of
\Xim (\Xip) candidates are shown in the \textit{left panel} (\textit{right panel}). Filled areas indicate the mass range of
the selected candidates. The vertical dashed black line shows the nominal PDG $\Xi$ mass. 
}
\label{fig:XiMass}
\end{figure*}

\Xim candidates were assembled by combining all \pim with those $\Lambda$ candidates having a reconstructed invariant mass within $\pm$15~\MeV of the nominal~\cite{Zyla:2020zbs} mass. A fitting procedure is applied using as parameters the decay position of the $V^{0}$ candidate, the momenta of both the $V^{0}$ decay tracks, and the momentum of the $\Xim$ daughter track. The z position of the $\Xim$ decay point is the intersection of the $\Lambda$ and \pim trajectories. The x and y coordinates of the $\Xi$ decay position are not subject to the minimization, as they are determined from the fitted parameters using momentum conservation. This procedure yields the decay
position and the momentum of the $\Xim$ candidate. 

Imposed cuts increase the significance of the \Xim signal. As the combinatorial background is largely due to particle production and decays close to the primary vertex, a distance of at least 12~cm was required between the primary and the \Xim vertices. Furthermore, the distance of the closest approach between the extrapolated \pim track from the \Xim decay and the primary vertex was required to be larger than 0.2~cm in the non-bending plane. To remove spurious \Xim candidates, their trajectory was required to have a distance of closest approach to the main vertex of less than 2~cm  (1~cm)  in the (non) bending plane.
The resulting $\Lambda$\pim invariant mass spectrum is shown in Fig.~\ref{fig:XiMass} (\textit{left}), where the \Xim peak is clearly visible. The \Xim candidates were selected within $\pm$15~\MeV of the nominal \Xim mass. Only events (95\%) with one \Xim candidate were retained. Precisely the same procedure was applied for the antiparticles, and the resulting \Xip peak is shown in Fig.~\ref{fig:XiMass} (\textit{right}).

\begin{figure*}[ht!]
\centering
\includegraphics[width=.49\textwidth]{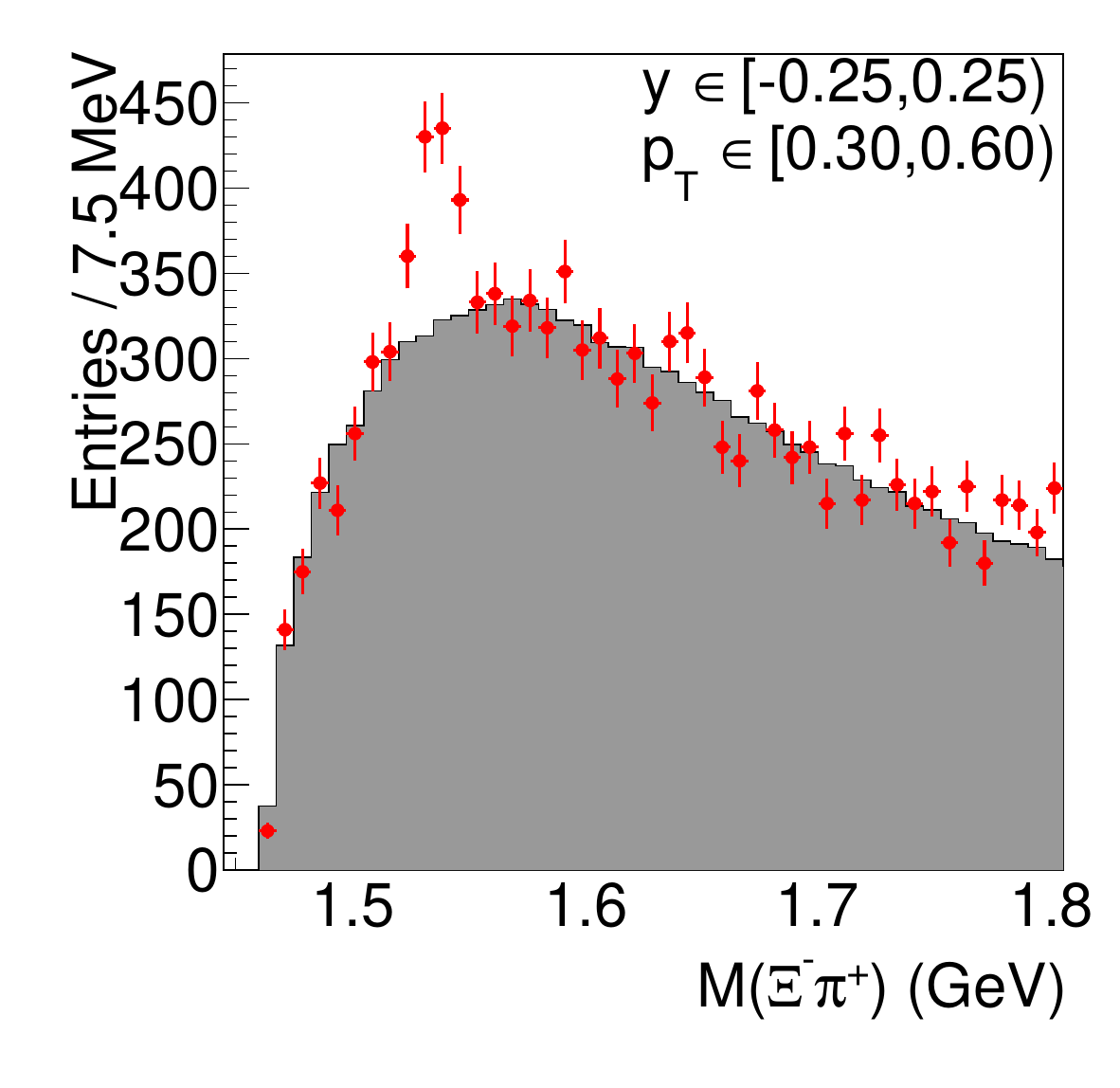}
\includegraphics[width=.49\textwidth]{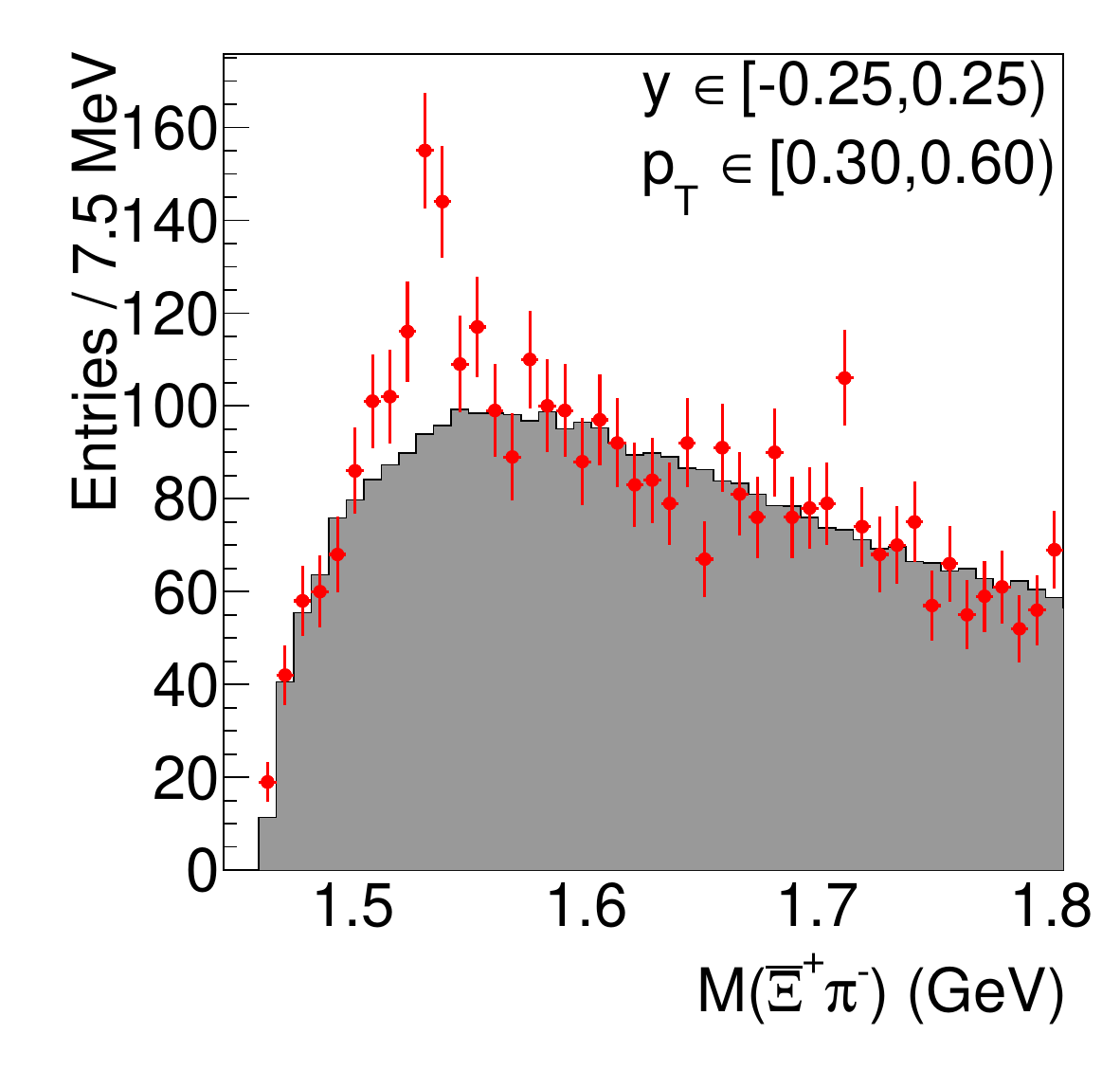}
\caption{(Color online) Invariant mass spectra of \Xim\pip and \Xip\pim combinations after applying all selection
criteria. The filled histograms are the normalized mixed-event background. 
}
\label{fig:massXir}
\end{figure*}

To search for the \Xir (\Xirb), the selected \Xim (\Xip) candidates were combined with primary \pip (\pim) tracks. To select pions originating from the primary vertex, their impact parameter $\lvert b_y \rvert$ was required to be less than 0.5~cm, and their \dedx to be within 3$\sigma$ of the nominal Bethe-Bloch value. 

\section{Signal extraction}\label{sec:signal}

For each $\Xir$ candidate, the invariant mass was calculated assuming the $\Xi$ and pion masses for the reconstructed candidate daughter particles and then histogrammed in \y,\pt bins. 
Examples of invariant mass distributions of \Xim\pip, \Xip\pim combinations are plotted in Fig.~\ref{fig:massXir}. The grey shaded histograms show the mixed-event background normalized to the number of real combinations, shown as red data points. The mixed background is determined by combining \Xim (\Xip) candidates with 1000 \pip (\pim) candidates from different events. The mixed background distributions were calculated for each y-\pt bin separately. The signal is determined by subtracting this normalized mixed-event background  (shaded histogram) from the experimental invariant mass spectrum. The background-subtracted signal was fitted to a Lorentzian function:
\begin{equation}
    \label{eq:lorenz}
    L(m)=\frac{1}{\pi}\frac{\frac{1}2{}\Gamma}{(m-m_{\Xir})^{2}+(\frac{1}{2}\Gamma)^{2}},
\end{equation}
where mass $m_\Xi$, width parameter $\Gamma$ and normalization constant are the fit parameters. The raw multiplicity of \Xir and \Xirb is calculated as the sum of the signal bins in a mass window whose width is defined as 3$\Gamma$ around the $\Xir$ mass of the fitted signal function (see.~Eq.~\ref{eq:lorenz}) to limit the propagation of statistical background fluctuations. 

\begin{figure*}[ht!]
\centering
\includegraphics[width=.49\textwidth]{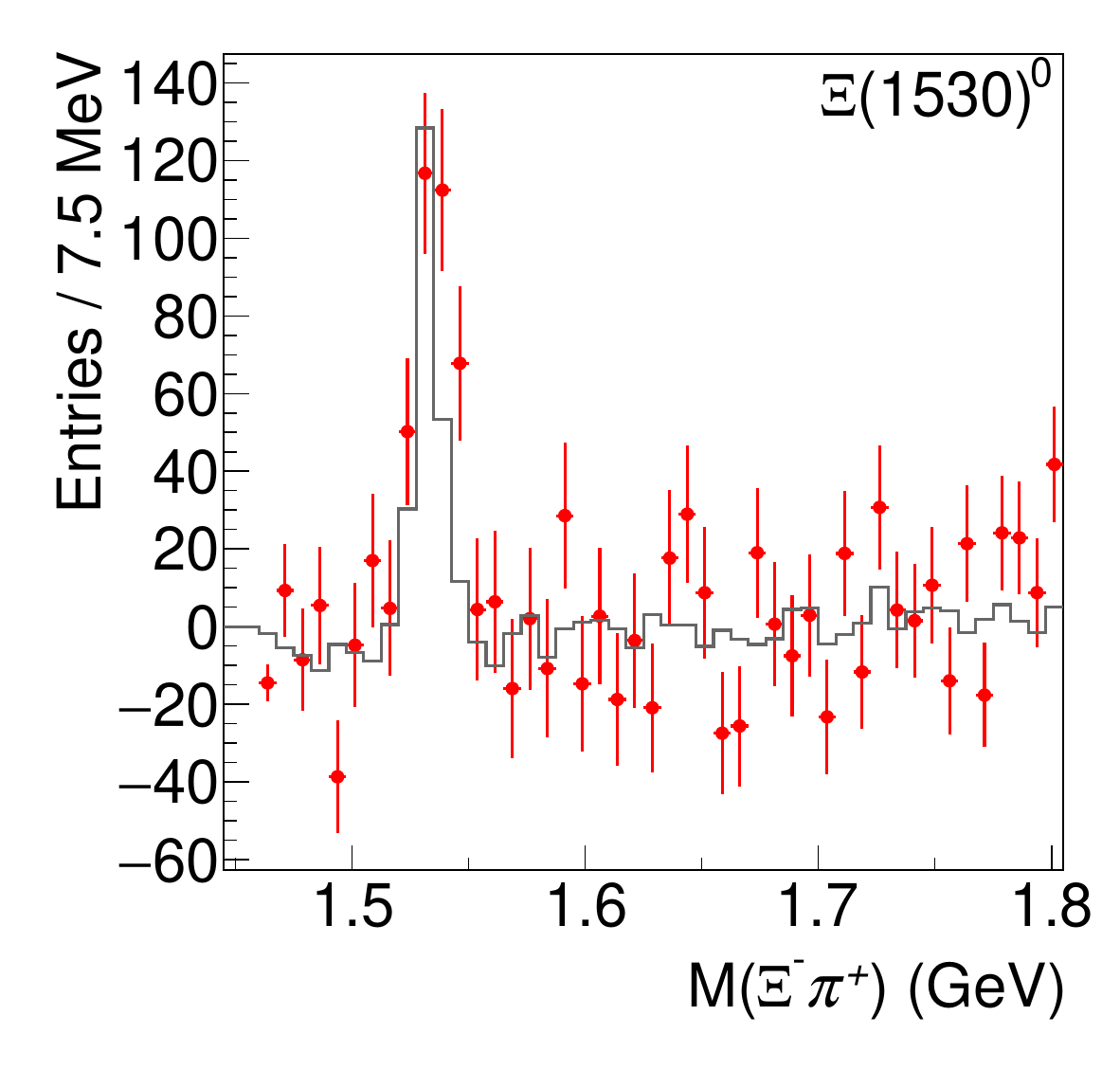}
\includegraphics[width=.49\textwidth]{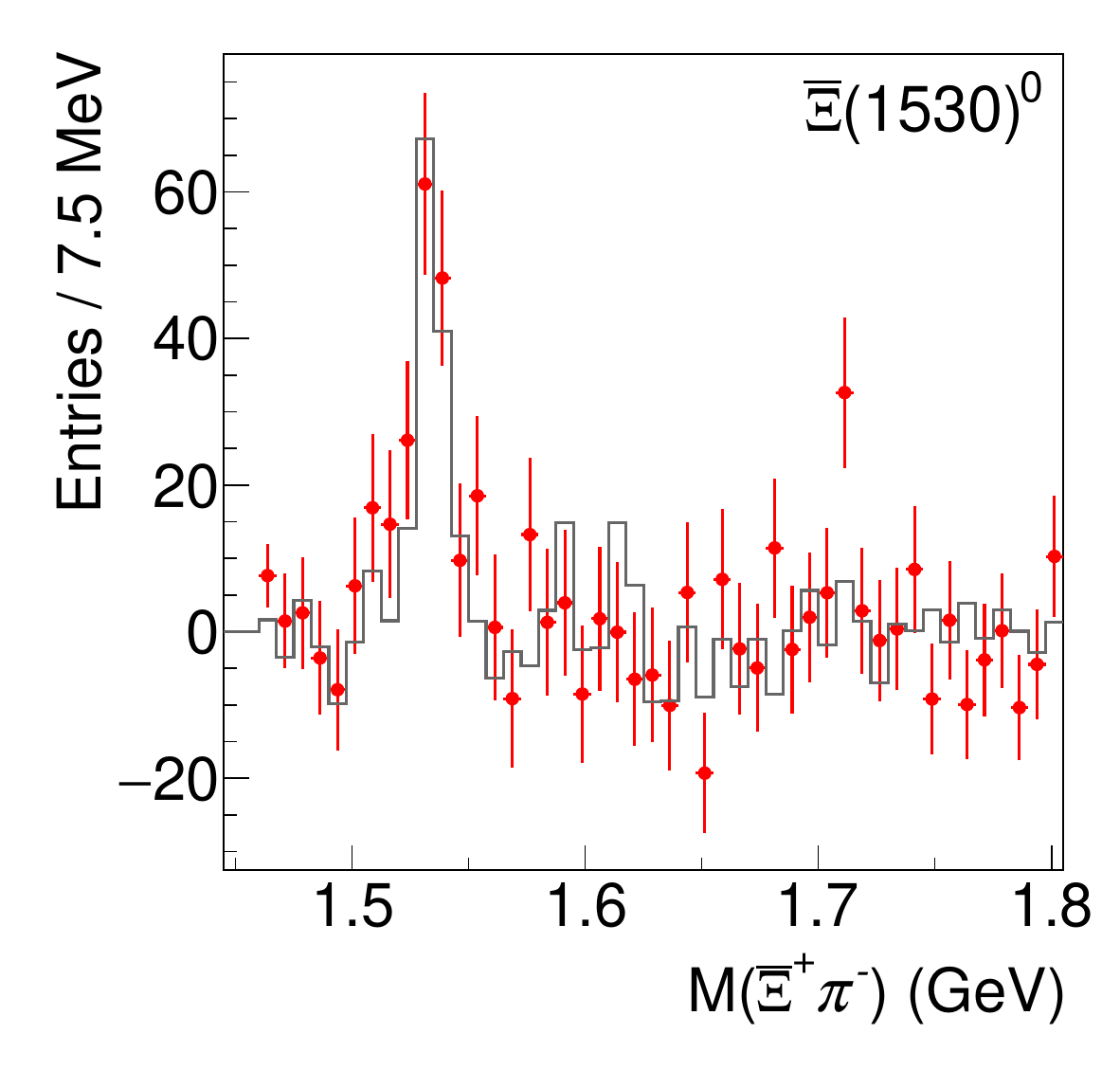}
\caption{(Color online)
Background-subtracted invariant mass distribution of \Xir (\textit{left} panel) and \Xirb (\textit{right} panel) in the rapidity range $-0.25 < y <0.25$ and transverse momentum range $0.30~\GeVc < \pt < 0.60~\GeVc$ from data (red points), and from \Epos~1.99 with full detector simulation and standard track and reconstruction procedures (gray histogram).
\label{fig:massdist}
}
\label{fig:massXiMC}
\end{figure*}

\begin{figure*}[ht!]
\centering
\includegraphics[width=.49\textwidth]{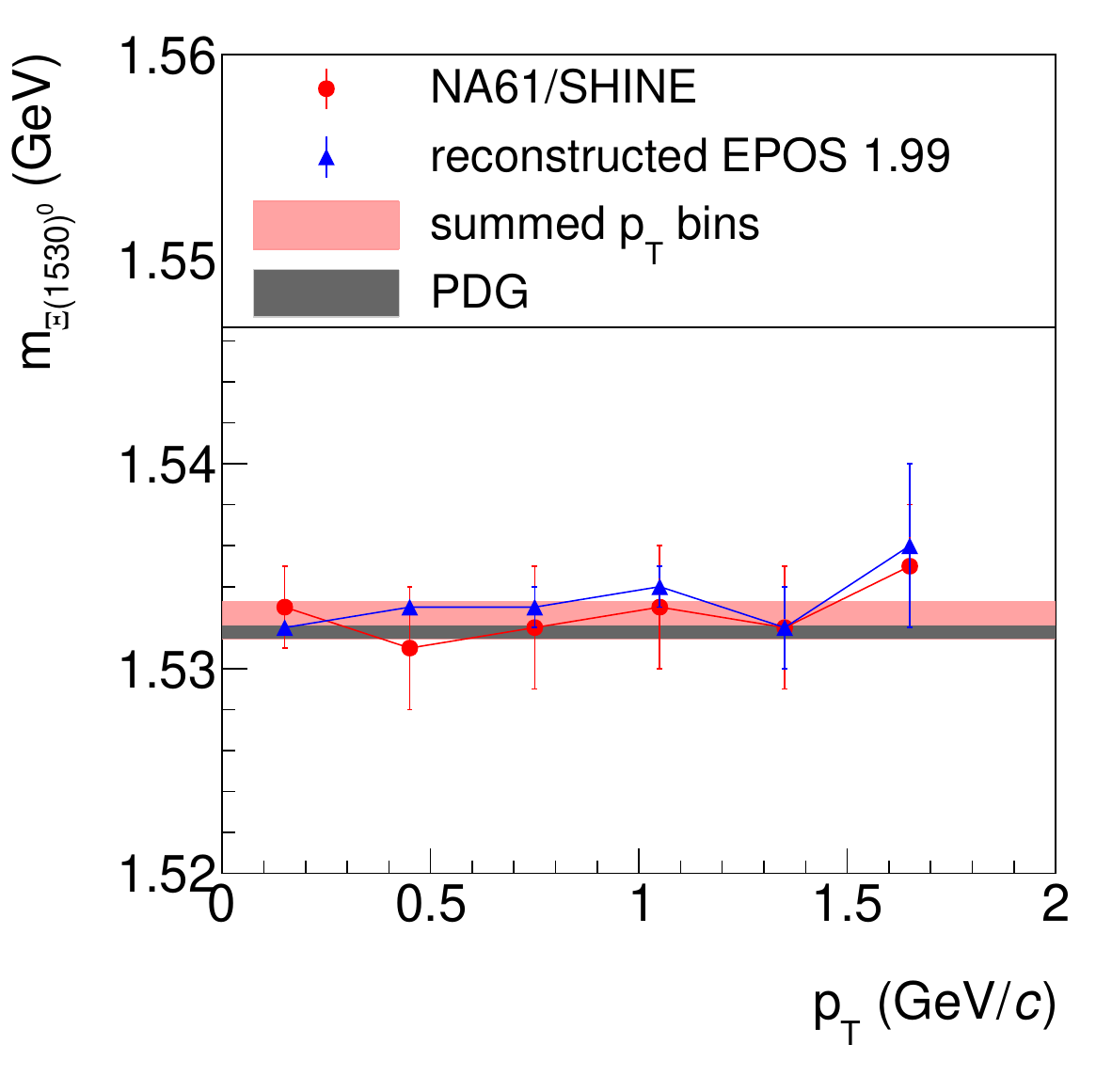}
\includegraphics[width=.49\textwidth]{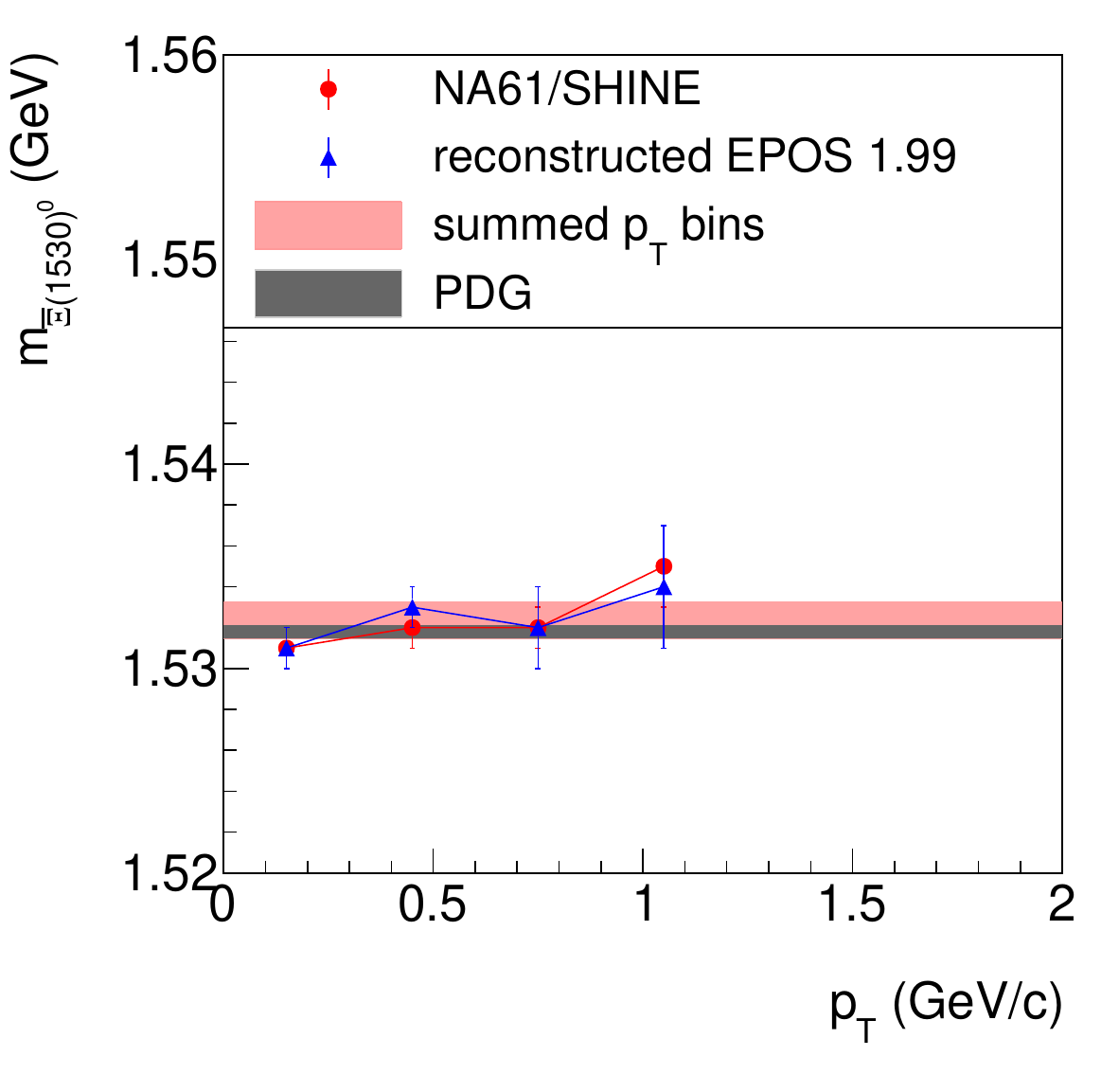}
\caption{(Color online) Mass of \Xir (left panel) and \Xirb (right panel) fitted at mid-rapidity $-0.25 < y <0.25$ as function of \pt for data (red points) and simulations (blue  points). The red horizontal band presents the uncertainty of the fitted mass of \Xir and \Xirb using the events of the full data set (combining all $y$, \pt bins) and the grey bands the PDG values~\cite{Zyla:2020zbs}.
}
\label{fig:massFit}
\end{figure*}

The fitted value of $\Gamma$ is larger than the PDG width due to the finite resolution of the detector. The width is close to expectations given by the analysis of inelastic \pp interactions generated by \Epos~1.99 with full detector simulation and standard track and $\Xir$ reconstruction procedures. The invariant mass distributions obtained experimentally and from simulations agree well, as shown for a selected \y,\pt bin as examples in Fig.~\ref{fig:massdist} for \Xir and \Xirb, respectively. The fitted mass value of $\Xir$ equals 1532.36 $\pm$ 0.94~\MeV and agrees within uncertainty with 1531.80 $\pm 0.32$~\MeV provided by PDG~\cite{Zyla:2020zbs}. More detailed verification of the stability of the fitted mass $m_\Xi$ was performed in the rapidity range $-0.25 < y <0.25$, as shown in Fig.~\ref{fig:massFit} where the masses are shown as a function of \pt. The fit results on $\Xir$ for both the real data and simulation are seen to agree with the PDG values.

\section{Corrections factors for yield determination}

A set of corrections was applied to the extracted raw results to determine the actual hyperons produced in inelastic \pp interactions. 

Interactions may contaminate the triggered and accepted events with the target vessel and other material in the target's vicinity. About 10\% of the data were collected without the liquid hydrogen in the target vessel to estimate the fraction of those events. After applying the event selection criteria described in Section~\ref{sec:method} and scaling to the same number of incoming beams, only 1\% of \Xir were found compared to the full target event sample. The correction was not applied for this contamination.

A detailed Monte-Carlo simulation is performed to quantify the losses due to acceptance limitations, detector inefficiencies, reconstruction shortcomings, analysis cuts, and re-interactions in the target. This simulation used complete events produced by the \Epos~1.99~\cite{Pierog:2009zt} event generator using a hydrogen target of appropriate length. The generated particles in each Monte-Carlo event are tracked through the detector using a GEANT3~\cite{Geant3} simulation of the \NASixtyOne apparatus. They are then reconstructed with the same software as used for real events. Numerous observables were confirmed to be similar to those of the data, such as residual distributions, widths of mass peaks, track multiplicities and their differential distributions, the number of events with no tracks in the detector, the cut variables and others. 

A correction factor is computed for each ($\y$, \pt) bin:
\begin{equation}
\label{eq:corr}
    C_{F} = n_{generated}^{MC}/n_{rec}^{MC},
\end{equation}
where $n_{rec}^{MC}$ is the number of reconstructed, selected, and identified $\Xir$s normalized to the number of analyzed events, and $n_{generated}^{MC}$ is the number of $\Xir$ generated by \Epos~1.99 normalized to the number of generated inelastic interactions. The raw multiplicity of $\Xir$ and $\Xirb$ are multiplied by $C_{F}$ to determine the true $\Xir$ and $\Xirb$ yields. These correction factors also include the branching fraction (66.7\%) of the \Xir and \Xirb decays into charged particles as well as the decay branching fraction of the $\Lambda$.

\section{Statistical and systematic uncertainties}\label{sec:systematics}

Statistical uncertainties of the yields receive contributions from the finite statistics of both the data and the correction factors derived from the simulations. The contribution from the statistical uncertainty of the data is much larger than that from the correction factors $C_{F}$. The statistical uncertainty of the ratio in Eq.\ref{eq:corr} was calculated assuming that the denominator $n_{rec}^{MC}$ is a subset of the nominator $n_{generated}^{MC}$ and thus has a binomial distribution.


Possible systematic uncertainties of final results (spectra and mean multiplicities) are due to the Monte Carlo procedure's imperfectness, e.g. the physics models and the detector response simulation - used to calculate the correction factors.

Several tests were performed to determine the magnitude of the different sources of possible systematic uncertainties:
\begin{enumerate}[(i)]
    \item Methods of event selection.
    
    Not all events which have tracks stemming from interactions of off-time beam particles are removed. A possible uncertainty due to this effect was estimated by changing by $\pm$1~$\mu s$ the width of the time window in which no second beam particle is allowed with respect to the nominal value of $\pm$2$~\mu$s. The maximum difference of the results was taken as an estimate of the uncertainty due to the selection. It was estimated to be 1-6\%.
	
	Another source of a possible bias are losses of inelastic events due to the interaction trigger. The S4 trigger selects mainly inelastic interactions and vetoes elastic scattering events. However, it will miss some of the inelastic events. To estimate the possible loss of $\Xi$s, simulations were done with and without the S4 trigger condition. The difference between these two results was taken as another contribution to the systematic uncertainty. The uncertainty due to the interaction trigger was calculated as half of the difference between these two results, which is 4-6\%. 
	
    The next source of systematic uncertainty related to the normalization came from the selection window for the z-position of the fitted vertex. To estimate the contribution of this systematic uncertainty, the selection criteria for the data and the \Epos~1.99 model were varied from $\pm$9~\cm to $\pm$10~\cm, and $\pm$11~\cm. The uncertainty due to the selection window for the z-position of the fitted vertex was estimated to be smaller than 2\%.
	
	\item Methods of $\Xir$ and $\Xirb$ candidates selection.
	To estimate the uncertainty related to the $\Xir$ and $\Xirb$ candidate selection, the following cut parameters were varied independently: 
	
	\begin{itemize}
	\item the distance cut between primary and decay vertex of \Xim (\Xip) was changed by $\pm$1~\cm and $\pm$2~\cm        yielding a possible bias of 1-6\%, 
	\item the extrapolated impact parameter of $\Xi$s in the $y$ direction at the main vertex z position  was changed       from 0.2~\cm to 0.1~\cm and 0.4~\cm, yielding a possible bias of up to 10\%,  
	\item the DCA of the pion ($\Xi$) daughter track to the main vertex was changed from 0.5~cm to 0.25 and 1~cm, yielding a possible bias of up to 9\%. 
    \end{itemize}
		
	\item Signal extraction.
	
	The uncertainty due to the signal extraction method was estimated by varying the invariant mass range used to determine the $\Xir$ yields by a change of $\pm$7~\MeV with respect to the nominal integration range and yielded a possible uncertainty up to 7\%.

	\end{enumerate}

The systematic uncertainty was calculated as the square root of the sum of squares of the described possible biases, assuming uncorrelated. The uncertainties are estimated for each ($\y$, \pt) bin separately.

\section{Experimental results}\label{sec:results}

This section presents results on inclusive $\Xir$ and $\Xirb$ production by strong interaction processes in inelastic \pp interactions at beam momentum of 158~\GeVc.

\subsection{Spectra and mean multiplicities}
\begin{figure*}[ht!]
\centering
\includegraphics[width=.48\textwidth]{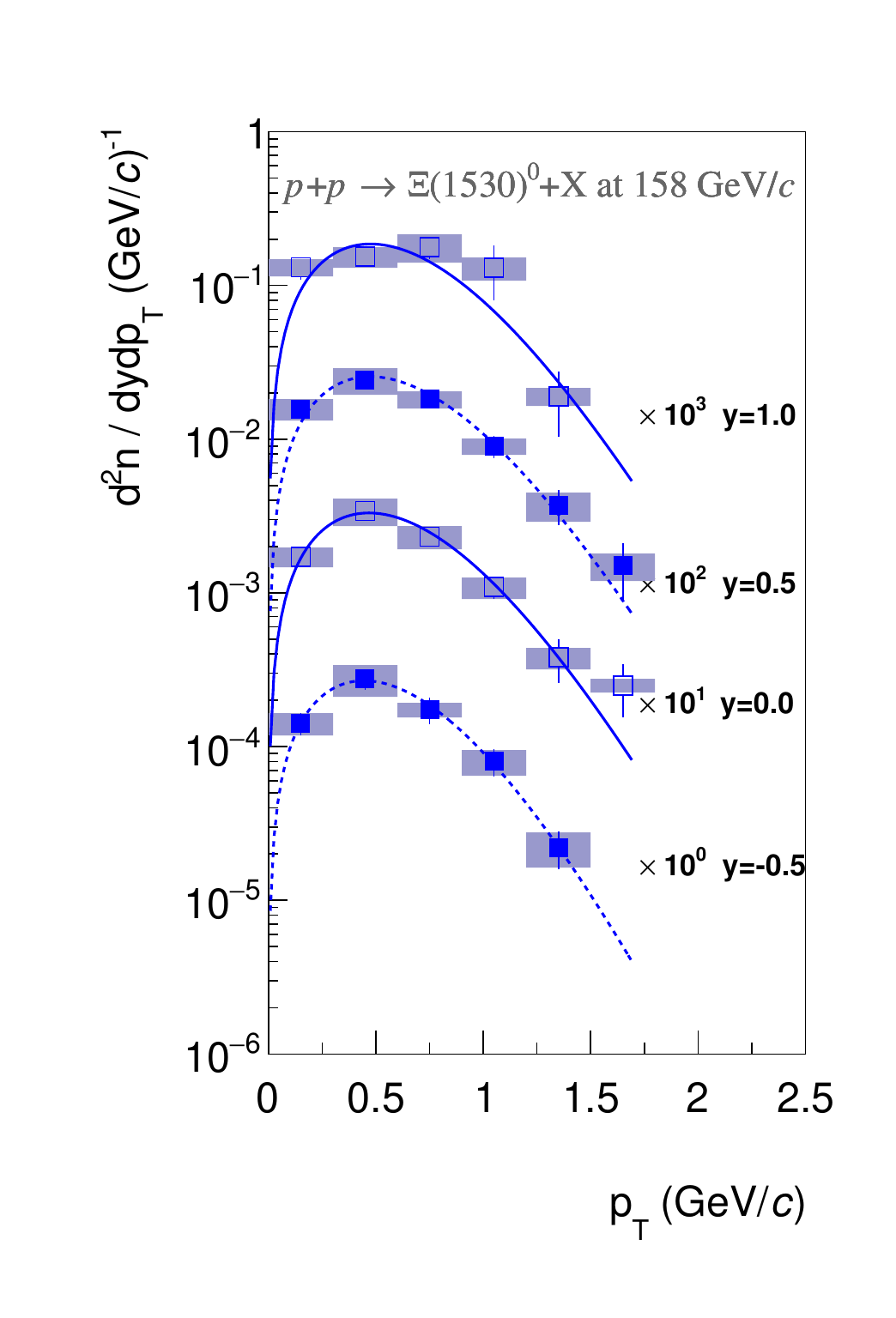}
\includegraphics[width=.48\textwidth]{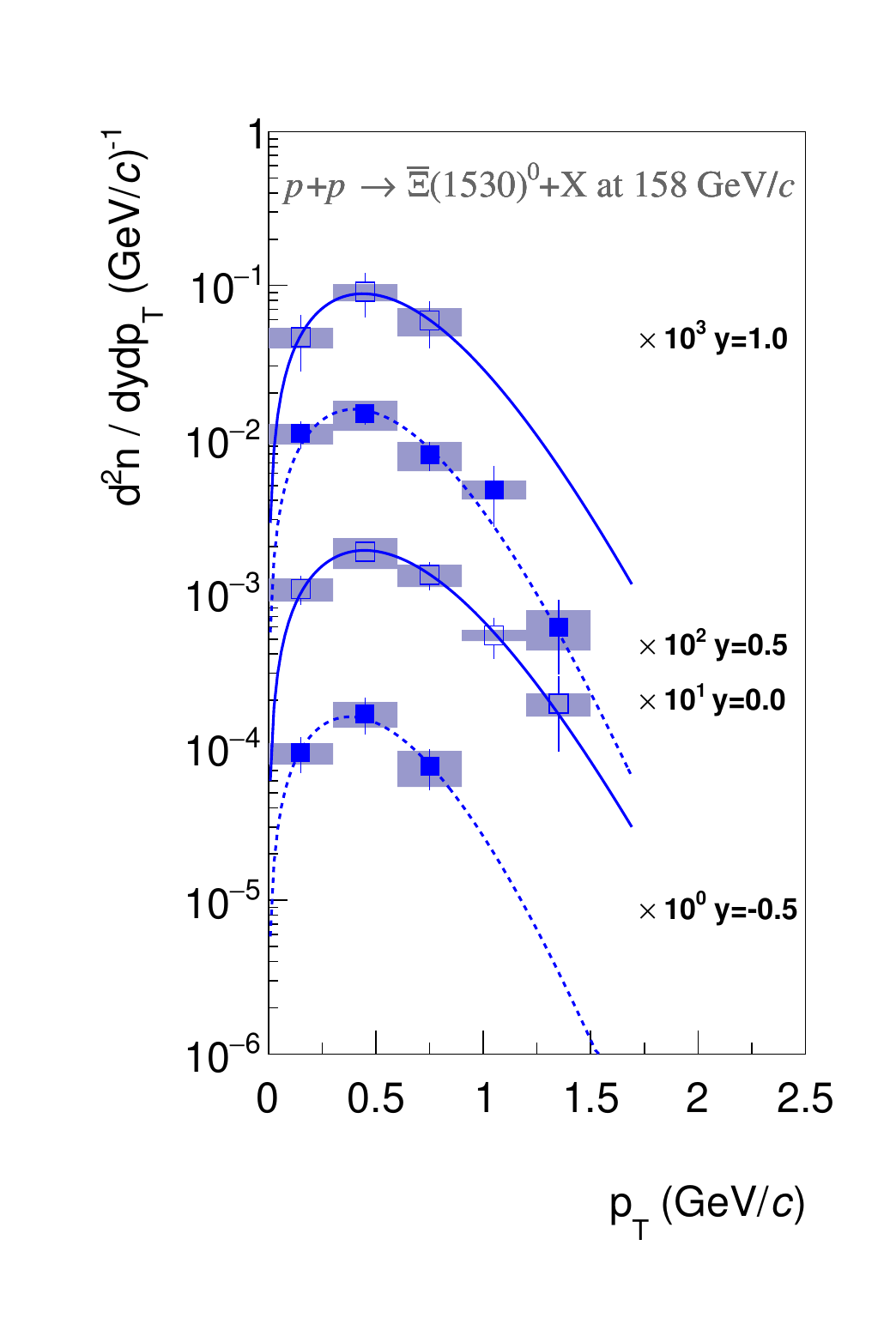}
\caption{(Color online) Transverse momentum spectra of $\Xir$ (\textit{left}) and $\Xirb$ (\textit{right}) in rapidity slices produced in inelastic \pp interactions at 158~\GeVc. Rapidity values given in the legends correspond to the middle of the corresponding interval. Statistical uncertainties are shown as vertical bars, and shaded bands show systematic uncertainties. Spectra are scaled for better visibility. Lines represent the fitted function (Eq.~{\ref{eq:inverse}}).}
\label{fig:results}
\end{figure*}

Double differential yields constitute the primary result of this paper. The $\Xir$ ($\Xirb$) yields are determined in 4 (4) rapidity and between 5 (3) and 6 (5) transverse momentum bins. The former is 0.5 units and the latter 0.3~\GeVc wide. The resulting ($\y$, \pt) yields are presented at the function of \pt in Fig.~\ref{fig:results}. The following exponential function can describe the transverse momentum spectra\cite{Hagedorn:1968jf,Broniowski:2004yh}:
\begin{equation}
 \frac{d^{2}n}{d\pt d\y}=\frac{S~c^{2}p_{T}}{T^{2} + m~T}\exp\left(-\frac{\mt - m}{T}\right),
\label{eq:inverse}
\end{equation}
where $m$ is the $\Xir$ mass, $m_{T}$ is the transverse mass defined as
$m_{T} = \sqrt{m^2 + (cp_{T})^2}$ and $c$ is the speed of light. The yields $S$ and the inverse slope parameters $T$ are determined by fitting the function to the data points in each rapidity bin.  The \pt spectra from successive rapidity intervals in Fig.~\ref{fig:results} are scaled for better visibility. Statistical uncertainties are shown as error bars, and shaded bands correspond to systematic uncertainties. Tables~\ref{tab:results:Xim} and \ref{tab:results:Xip} list the numerical values of the results shown in Fig.~\ref{fig:results}. The resulting inverse slope parameters are listed in Table~\ref{tab:dndy}. 

\begin{table}[ht]
\caption{Numerical values of double-differential spectra of $\Xir$ produced in inelastic \pp interactions at 158~\GeVc beam momentum. Rapidity and transverse momentum values correspond to the middle of the presented bin. The first value is the particle multiplicity, the second represents the statistical uncertainty, and the third corresponds to
the estimated systematic uncertainty.}\label{tab:results:Xim}
\centering
\small
\begin{tabular}{|c|c|c|c|c|}
\multicolumn{5}{c}{}\\
\multicolumn{5}{c}{\large{$\Xir$: $\frac{d^{2}n}{d\y d\pt}$ $\left(\GeVc\right)^{-1}$}} \\ 
\multicolumn{5}{c}{}\\
\hline
&&&&\\
\pt (\GeVc) & $\y\approx$ -0.5 & $\y\approx$ 0.0 & $\y\approx$ 0.5 & $\y\approx$ 1.0\\
&&&&\\
\hline
0.15 & (1.42$\pm$0.23$\pm$0.24)$\times10^{-4}$ & (1.72$\pm$0.20$\pm$0.24)$\times10^{-4}$ & (1.57$\pm$0.18$\pm$0.25)$\times10^{-4}$ & (1.31$\pm$0.21$\pm$0.16)$\times10^{-4}$ \\
0.45 & (2.75$\pm$0.41$\pm$0.49)$\times10^{-4}$ & (3.41$\pm$0.31$\pm$0.68)$\times10^{-4}$ & (2.43$\pm$0.21$\pm$0.47)$\times10^{-4}$ & (1.54$\pm$0.21$\pm$0.24)$\times10^{-4}$ \\
0.75 & (1.74$\pm$0.16$\pm$0.15)$\times10^{-4}$ & (2.32$\pm$0.28$\pm$0.40)$\times10^{-4}$ & (1.82$\pm$0.20$\pm$0.24)$\times10^{-4}$ & (1.78$\pm$0.29$\pm$0.37)$\times10^{-4}$ \\
1.05 & (0.80$\pm$0.19$\pm$0.18)$\times10^{-4}$ & (1.09$\pm$0.18$\pm$0.17)$\times10^{-4}$ & (0.90$\pm$0.15$\pm$0.10)$\times10^{-4}$ & (1.30$\pm$0.50$\pm$0.22)$\times10^{-4}$ \\
1.35 & (0.22$\pm$0.06$\pm$0.06)$\times10^{-4}$ & (0.38$\pm$0.12$\pm$0.06)$\times10^{-4}$ & (0.37$\pm$0.09$\pm$0.08)$\times10^{-4}$ & (0.19$\pm$0.08$\pm$0.03)$\times10^{-4}$ \\
1.65 & -                                       & (0.25$\pm$0.09$\pm$0.03)$\times10^{-4}$ & (0.15$\pm$0.06$\pm$0.03)$\times10^{-4}$ & - \\
\hline
\end{tabular}
\end{table}

\begin{table}[ht]
\caption{Numerical values of double-differential spectra of $\Xirb$ produced in inelastic \pp interactions at 158~\GeVc beam momentum. Rapidity and transverse momentum values correspond to the middle of the presented bin. The first value is the particle multiplicity, the second represents the statistical uncertainty, and the third corresponds to
the estimated systematic uncertainty.}\label{tab:results:Xip}
\centering
\small
\begin{tabular}{|c|c|c|c|c|}
\multicolumn{5}{c}{}\\
\multicolumn{5}{c}{\large{$\Xirb$: $\frac{d^{2}n}{d\y d\pt}$ $\left(\GeVc\right)^{-1}$}} \\ 
\multicolumn{5}{c}{}\\
\hline
&&&&\\
\pt (\GeVc) & $\y\approx$ -0.5 & $\y\approx$ 0.0 & $\y\approx$ 0.5 & $\y\approx$ 1.0\\
&&&&\\
\hline
0.15 &  (0.91$\pm$0.24$\pm$0.14)$\times10^{-4}$ & (1.06$\pm$0.23$\pm$0.18)$\times10^{-4}$ & (1.09$\pm$0.22$\pm$0.17)$\times10^{-4}$ & (0.46$\pm$0.18$\pm$0.07)$\times10^{-4}$  \\
0.45 &  (1.64$\pm$0.45$\pm$0.31)$\times10^{-4}$ & (1.85$\pm$0.26$\pm$0.42)$\times10^{-4}$ & (1.46$\pm$0.22$\pm$0.32)$\times10^{-4}$ & (0.91$\pm$0.29$\pm$0.11)$\times10^{-4}$ \\
0.75 &  (0.74$\pm$0.22$\pm$0.19)$\times10^{-4}$ & (1.31$\pm$0.27$\pm$0.22)$\times10^{-4}$ & (0.79$\pm$0.17$\pm$0.17)$\times10^{-4}$ & (0.59$\pm$0.20$\pm$0.13)$\times10^{-4}$ \\
1.05 & -                                        & (0.53$\pm$0.16$\pm$0.06)$\times10^{-4}$ & (0.47$\pm$0.20$\pm$0.08)$\times10^{-4}$ & -   \\
1.35 & -                                        & (0.19$\pm$0.09$\pm$0.03)$\times10^{-4}$ & (0.06$\pm$0.03$\pm$0.02)$\times10^{-4}$ & -\\
\hline
\end{tabular}
\end{table}

\begin{figure}[!ht]
                \begin{center}
                \includegraphics[width=0.7\textwidth]{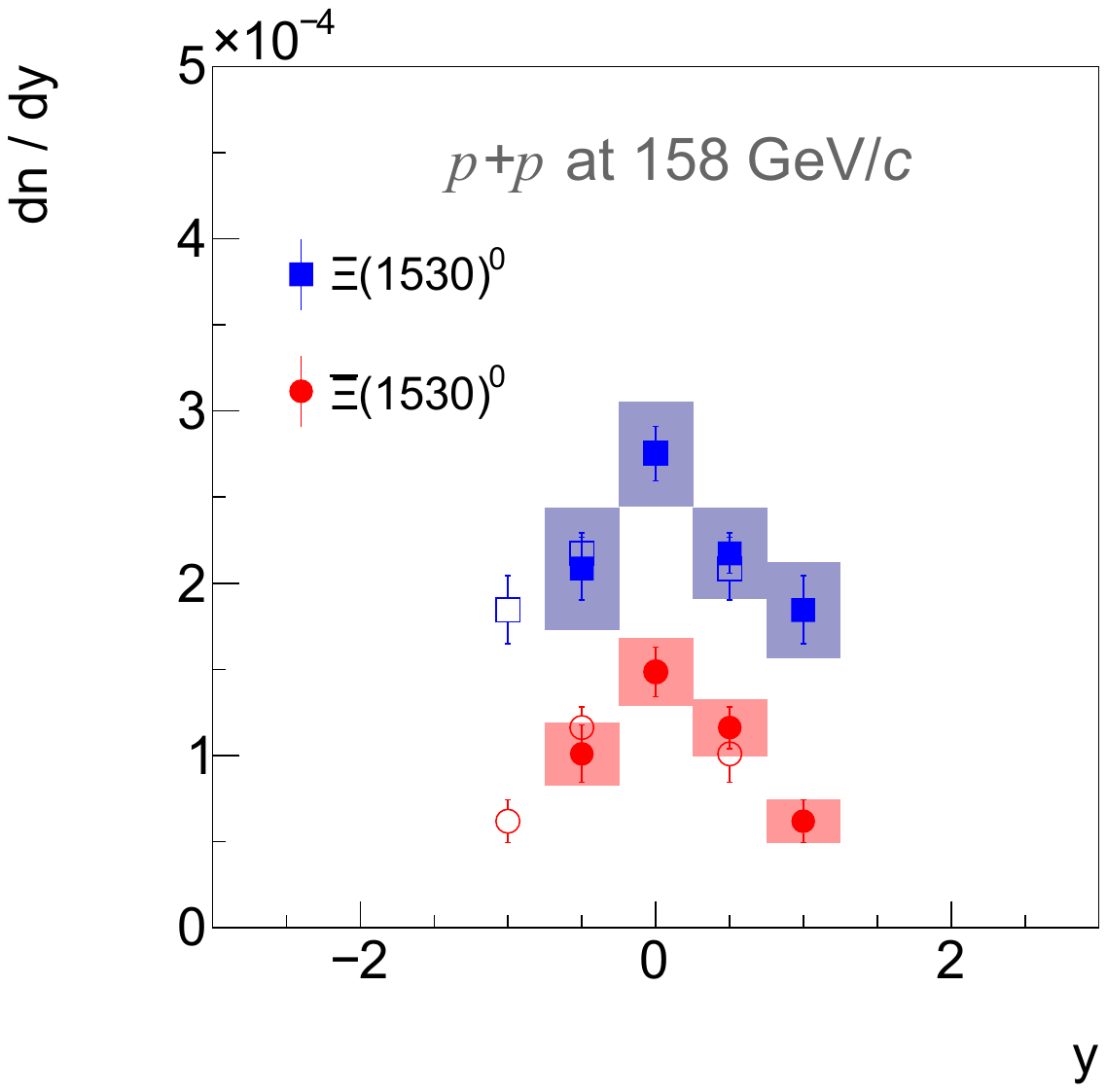}
                \end{center}
                \caption{(Color online) Rapidity spectra of $\Xir$ (blue squares) and $\Xirb$ (red circles) produced in inelastic \pp interactions at 158~\GeVc. Statistical uncertainties are shown by vertical bars, and shaded bands correspond to systematic uncertainties of the measurements. Full symbols show measured points, open symbols values reflected around mid-rapidity.} 
                \label{fig:finaldndy}
        \end{figure}

The yields as function of rapidity were then obtained by summing the measured transverse momentum spectra and extrapolating them into the unmeasured regions using the fitted functions given by Eq.~\ref{eq:inverse}. The resulting rapidity distributions are shown in Fig.~\ref{fig:finaldndy}. The statistical uncertainties are shown as error bars. They were calculated as the square root of the sum of the squares of the statistical uncertainties of the contributing bins. The systematic uncertainties (shaded bands) were calculated as the square root of squares of systematic uncertainty as described in Sec.~\ref{sec:systematics} and half of the extrapolated yield. The numerical values of the rapidity distribution $dN/dy$ and their errors are listed in Table~\ref{tab:dndy}.

\begin{table}[ht]
\caption{Numerical values of rapidity spectra of $\Xir$ and $\Xirb$ produced in inelastic \pp interactions at 158~\GeVc beam momentum and fitted inverse slope parameter T (see eq.~\ref{eq:inverse}). Rapidity values correspond to the middle of the presented bin. The first value is the particle multiplicity, the second represents the statistical uncertainty, and the third corresponds to
the estimated systematic uncertainty.}\label{tab:dndy}
\centering
\small
\begin{tabular}{|c|c|c||c|c|}
\multicolumn{5}{c}{}\\
\hline
&&&&\\
$\y$   &   $\Xir$: $\frac{dn}{d\y}$ &  $\Xir$: T (\MeV)               &   $\Xirb$:  $\frac{dn}{d\y}$  &  $\Xirb$: T (\MeV)\\
&&&&\\
\hline

-0.5 & (2.08$\pm$0.18$\pm$0.35)$\times10^{-4}$ & 124$\pm$10$\pm$12 & (1.01$\pm$0.17$\pm$0.18)$\times10^{-4}$ &  91$\pm$21$\pm$16\\
0.0  & (2.75$\pm$0.16$\pm$0.30)$\times10^{-4}$ & 136$\pm$10$\pm$12 & (1.48$\pm$0.14$\pm$0.19)$\times10^{-4}$ & 125$\pm$15$\pm$12\\
0.5  & (2.17$\pm$0.12$\pm$0.26)$\times10^{-4}$ & 141$\pm$11$\pm$13 & (1.15$\pm$0.12$\pm$0.16)$\times10^{-4}$ &  97$\pm$19$\pm$14\\
1.0  & (1.84$\pm$0.20$\pm$0.28)$\times10^{-4}$ & 143$\pm$17$\pm$14 & (0.62$\pm$0.12$\pm$0.12)$\times10^{-4}$ & 121$\pm$49$\pm$24\\
\hline
\end{tabular}
\end{table}


A correction factor based on the \Epos model is calculated and used to extrapolate into the unmeasured regions. The correction factor is defined as the ratio of the \Xir (\Xirb) multiplicity from the \Epos model in the measurement region to the total \Epos \Xir (\Xirb) multiplicity and equals 1.52 (1.27). Summing the data points and multiplying by the obtained correction factor allows to obtain the mean multiplicities $\Xir =$ (6.73 $\pm$ 0.25 $\pm$ 0.67)$\times10^{-4}$ and $\Xirb = $ (2.71 $\pm$ 0.18 $\pm$ 0.18)$\times10^{-4}$. In addition a systematic error of 50\% of the extrapolated yield was included.


 The rapidity densities ($dn/d\y$) at mid-rapidity of $\Xir$ and $\Xirb$ produced in inelastic \pp interactions at $\sqrt{s_{NN}}=17.3$~\GeV collisions can be compared with results from ALICE at CERN LHC measured at $\sqrt{s_{NN}}$ = 7~\TeV~\cite{Abelev:2014qqa}. The ratios of $\Xi(1530)$ to $\Xi$~\cite{Abelev:2012jp, Aduszkiewicz:2020dyw} at mid-rapidity in \pp interactions at 17.3~\GeV and 7~\TeV are shown in Table~\ref{tab:ratioToXi}. The ratio of $\Xi(1530)$ to $\Xi$ at these two energies are similar: 0.294 $\pm$ 0.017 $\pm$ 0.047 at 17.3~\GeV and 0.3241 $\pm$ 0.0053 $^{+0.0325}_{0.0275}$ at 7~\TeV, while the yields of \Xir increase with collision energy by an order of magnitude from (2.75$\pm$0.18$\pm$0.58)$\times10^{-4}$ at 17.3~\GeV to (2.56$\pm$0.07$^{+0.40}_{-0.37}$)$\times10^{-3}$ at 7~\TeV.

\begin{table}[ht]
\caption{The ratio of \Xir to \Xim, of \Xirb to \Xip, and of (\Xir+\Xirb) to (\Xim+\Xip) in mid-rapidity in \pp interactions at 17.3~\GeV~\cite{Aduszkiewicz:2020dyw} and 7~\TeV~\cite{Abelev:2014qqa,Abelev:2012jp}. Systematic uncertainties were calculated with the assumption that the uncertainties of $\Xi$ and $\Xi(1530)$ are independent.}\label{tab:ratioToXi}
\centering
\small
\begin{tabular}{|c|c|c|}
\hline
&&\\
 & $\sqrt{s_{NN}}$ = 17.3~\GeV & $\sqrt{s_{NN}}$ = 7~\TeV \\
&&\\
\hline
\Xir / \Xim  & 0.267 $\pm$ 0.018 $\pm$ 0.058  & \\
&&\\
\Xirb / \Xip & 0.364 $\pm$ 0.040 $\pm$ 0.078  & \\
&&\\
$\frac{\Xir + \Xirb}{\Xim + \Xip}$ & 0.294 $\pm$ 0.017 $\pm$ 0.047 & 0.3241 $\pm$ 0.0053 $^{+0.0325}_{-0.0275}$\\
\hline

\end{tabular}


\end{table}


\subsection{Anti–baryon/baryon ratios}
The \NASixtyOne measurement of $\Xir$ hyperon production allows to determine anti–baryon/baryon ratios at central rapidity and ratios of total mean multiplicities in \pp collisions. The systematic uncertainties of \Xir and \Xirb are correlated. Therefore the systematic uncertainty of $\Xirb/\Xir$ had to be determined separately (following the procedure described in Sec.~\ref{sec:systematics}). The $\Xirb/\Xir$ ratios as a function of rapidity and transverse momentum are listed in Table~\ref{tab:resultsratio}. The ratio of the rapidity spectra are listed in Table~\ref{tab:dndyratio} and drawn in Fig.~\ref{fig:dnmodel}(c).  
The small value of the  ratio of mean multiplicities $\left\langle \Xirb \right\rangle / \left\langle \Xir \right\rangle = $  0.40 $\pm$ 0.03 $\pm$ 0.05 emphasizes the strong suppression of $\Xirb$ production at CERN SPS energies. This effect disappears, as expected, at the much higher LHC energies~\cite{Abelev:2014qqa}.

\begin{table}[ht]
\caption{The $\Xirb/\Xir$ ratio in inelastic \pp interactions at 158~\GeVc beam momentum. Rapidity and transverse momentum values correspond to the middle of the presented bin. The first value is the particle ratio, the second represents the statistical uncertainty, and the  third corresponds to
the estimated systematic uncertainty.}\label{tab:resultsratio}
\centering
\small
\begin{tabular}{|c|c|c|c|c|}
\multicolumn{5}{c}{}\\
\multicolumn{5}{c}{\large{$\Xirb/\Xir$}} \\ 
\multicolumn{5}{c}{} \\ 
\hline
&&&&\\
\pt (\GeVc) & $\y\approx$ -0.5 & $\y\approx$ 0.0 & $\y\approx$ 0.5 & $\y\approx$ 1.0\\
&&&&\\
\hline
0.15 & 0.64$\pm$0.020$\pm$0.15 & 0.62$\pm$0.15$\pm$0.08 & 0.69$\pm$0.16$\pm$0.10 & 0.35$\pm$0.15$\pm$0.05\\
0.45 & 0.60$\pm$0.019$\pm$0.18 & 0.54$\pm$0.09$\pm$0.08 & 0.60$\pm$0.10$\pm$0.08 & 0.59$\pm$0.20$\pm$0.08\\
0.75 & 0.42$\pm$0.015$\pm$0.12 & 0.56$\pm$0.14$\pm$0.07 & 0.43$\pm$0.10$\pm$0.07 & 0.33$\pm$0.12$\pm$0.05\\
1.05 & -                       & 0.48$\pm$0.19$\pm$0.07 & 0.52$\pm$0.24$\pm$0.08 & - \\
1.35 & -                       & 0.50$\pm$0.30$\pm$0.06 & -                      & - \\

\hline
\end{tabular}
\end{table}

\begin{table}[ht]
\caption{Ratio of \pt integrated yields versus rapidity of $\Xirb$ and $\Xir$ produced in inelastic \pp interactions at 158~\GeVc beam momentum. Rapidity values correspond to the middle of the presented bin. The first value is the particle multiplicity, the second represents the statistical uncertainty, and the  third corresponds to
the estimated systematic uncertainty.}\label{tab:dndyratio}
\centering
\small
\begin{tabular}{|c|c|}
\multicolumn{2}{c}{}\\
\hline
&\\
$\y$   &   $\Xirb/\Xir$ \\
&\\
\hline
-0.5 & 0.48 $\pm$ 0.10 $\pm$ 0.07  \\
0.0  & 0.54 $\pm$ 0.06 $\pm$ 0.07 \\
0.5  & 0.53 $\pm$ 0.06 $\pm$ 0.07 \\
1.0  & 0.33 $\pm$ 0.11 $\pm$ 0.04 \\
\hline
\end{tabular}
\end{table}


\section{Comparison with models}\label{sec:comparison}

The new \NASixtyOne measurements of $\Xir$ and \Xirb production 
are essential for understanding multi-strange particle production in elementary hadron interactions. 

Measurements of multi-strange hyperon production at intermediate energies possibly provide new insight into string formation and decay. In the string picture, \pp collisions create string "excitations", which are hypothetical objects that decay into hadrons according to longitudinal phase space. Multi-strange baryons and their anti-particles play a special role in this context. Their pairwise production in string decays will increase the anti-strange-baryon to the strange-baryon ratio significantly compared to the anti-baryon to baryon ratio.
Indications of this expectation are observed in the $\Xi$ and $\overline{\Xi}$ yields obtained from the \Urqmd model shown below.

\begin{figure*}[ht!]
\centering
\includegraphics[width=.49\textwidth]{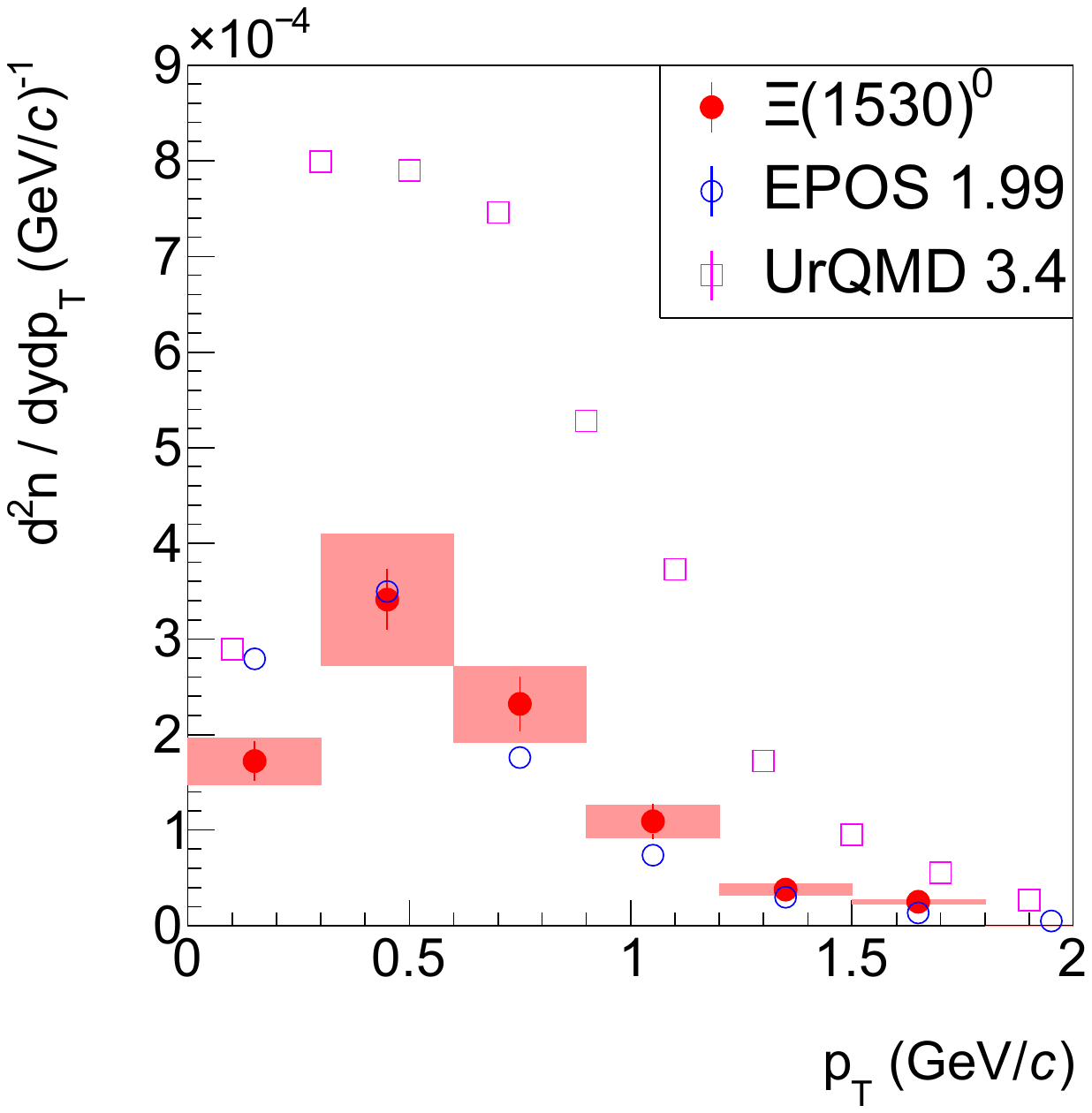}
\includegraphics[width=.49\textwidth]{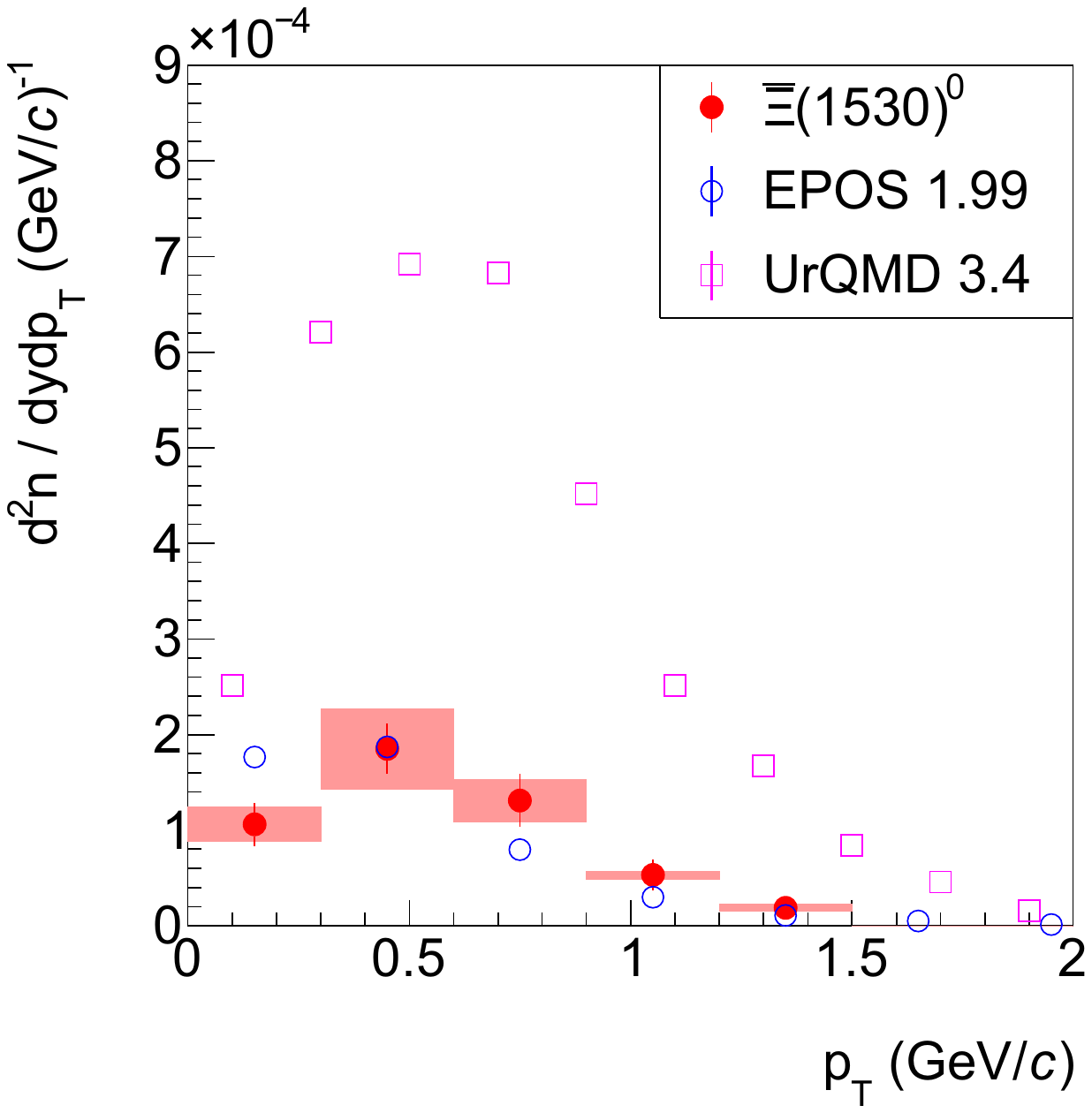}
\caption{(Color online) Transverse momentum spectra at mid-rapidity of $\Xir$ (\textit{left}) and $\Xirb$ (\textit{right}) produced in inelastic \pp interactions at 158~\GeVc. Shaded bands show systematic uncertainties. \Urqmd 3.4~\cite{Bass:1998ca,Bleicher:1999xi} and \Epos~1.99~\cite{Werner:2008zza} 
predictions are shown as magenta and blue markers, respectively.}
\label{fig:ptmodel}
\end{figure*}

\begin{figure*}[ht!]
\centering
\includegraphics[width=.325\textwidth]{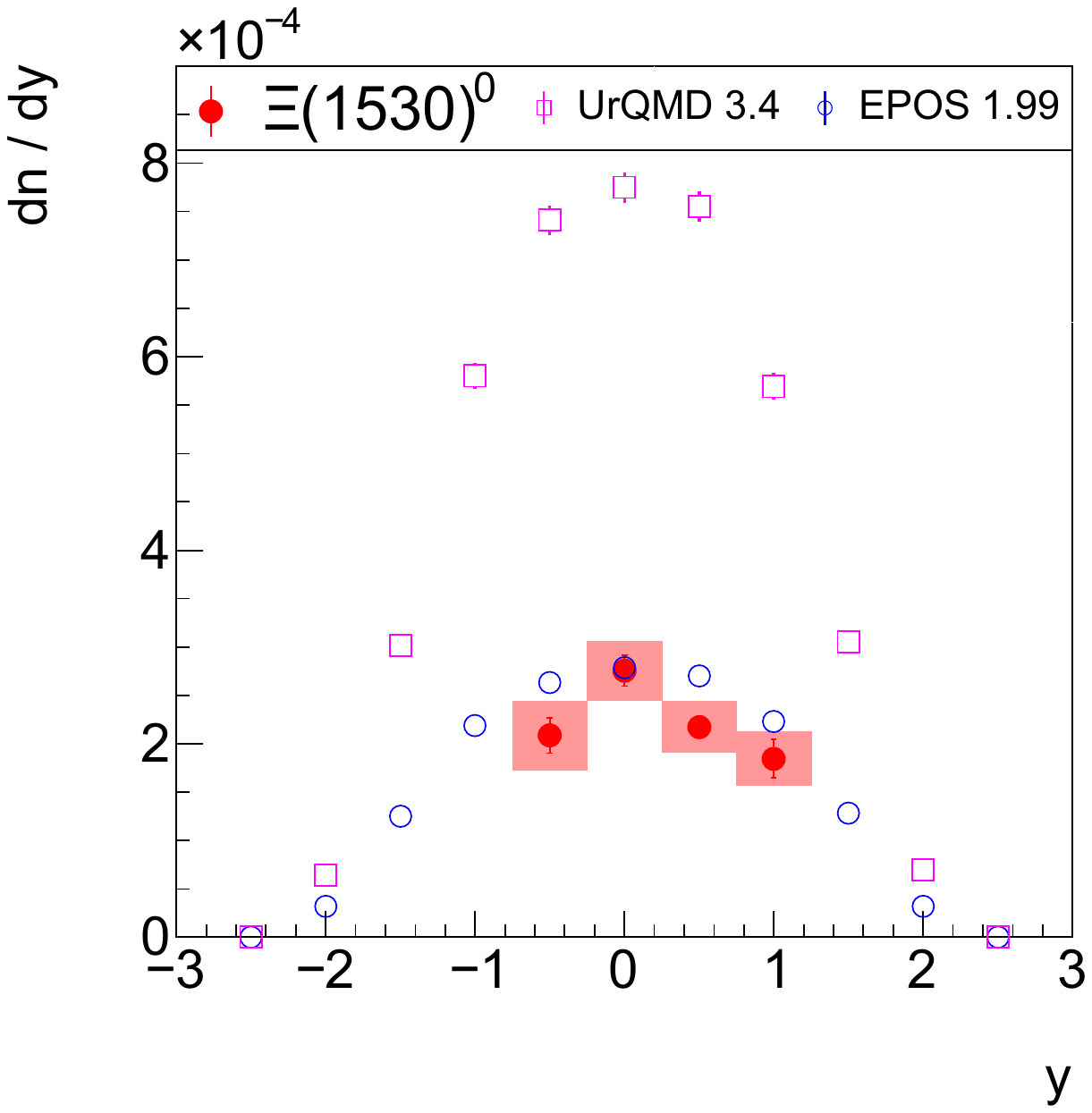}
\includegraphics[width=.325\textwidth]{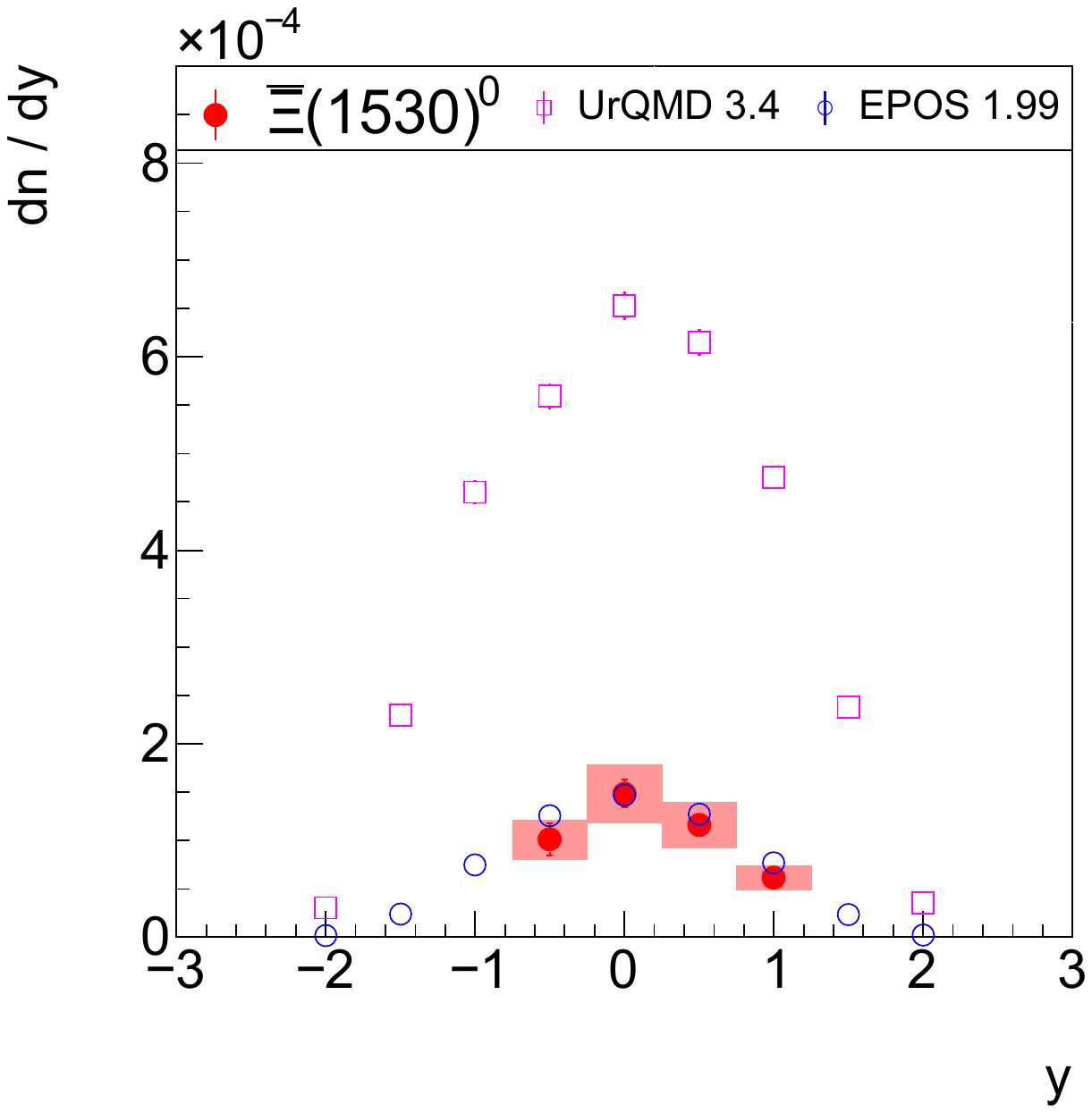}
\includegraphics[width=.325\textwidth]{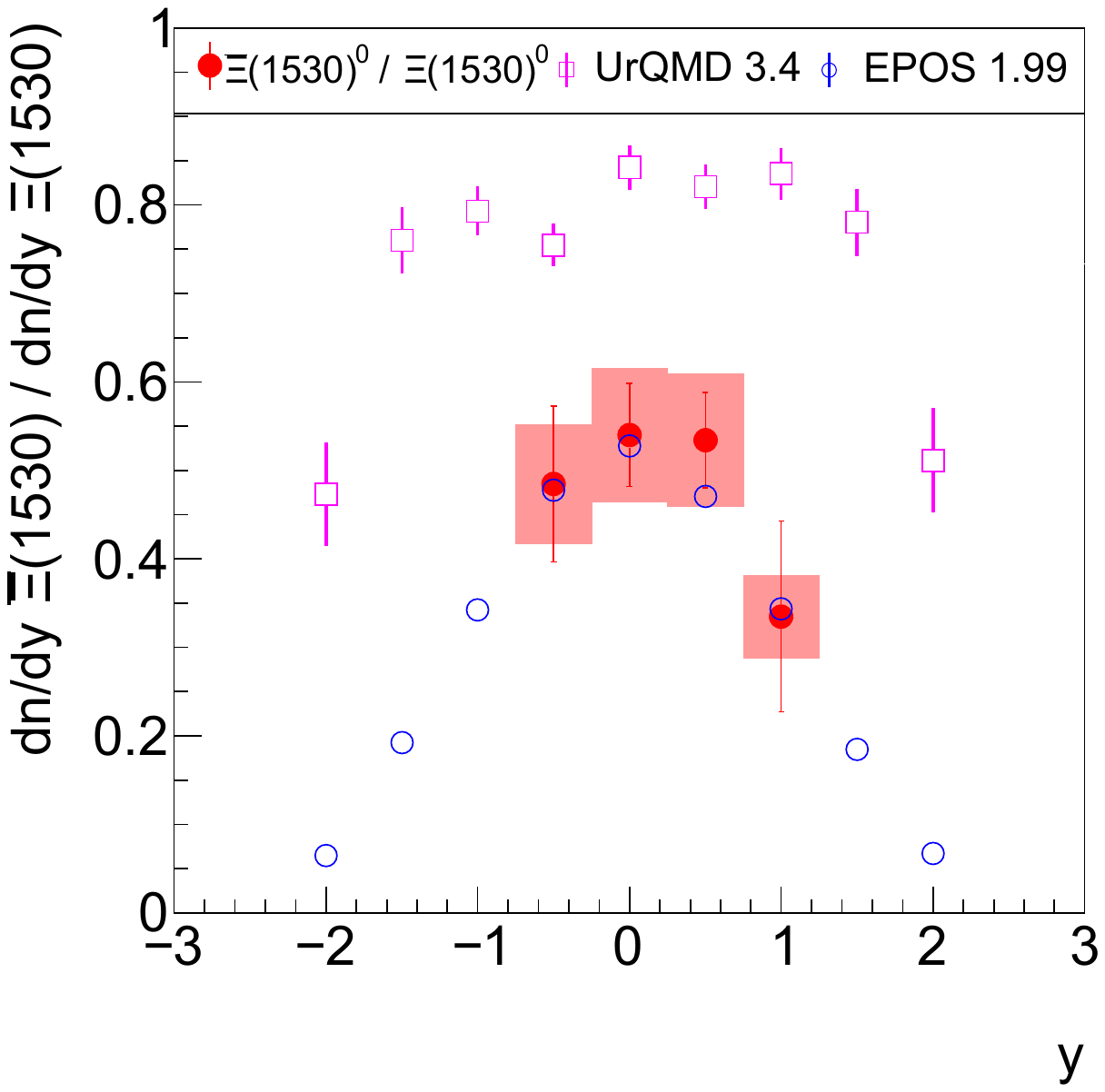}
\caption{(Color online) Rapidity spectra of $\Xir$ (\textit{left}), $\Xirb$ (\textit{middle}) and $\Xirb/\Xir$ ratio (\textit{right}) measured in inelastic \pp interactions at 158~\GeVc. Shaded bands show systematic uncertainties. \Urqmd 3.4~\cite{Bass:1998ca,Bleicher:1999xi} and \Epos~1.99~\cite{Werner:2008zza}
predictions are shown as magenta and blue
points, respectively.}
\label{fig:dnmodel}
\end{figure*}

The experimental results of \NASixtyOne are compared with predictions of the \Epos 1.99~\cite{Werner:2008zza} and \Urqmd~3.4~\cite{Bass:1998ca,Bleicher:1999xi} models. 
In \Epos, the reaction proceeds from the excitation of strings according to Gribov-Regge theory to string fragmentation into hadrons. \Urqmd starts with a hadron cascade based on elementary cross sections for the production of states, which either decay (mostly at low energies) or are converted into strings that fragment into hadrons (mostly at high energies). These are the only models that provide the history of the produced particle needed to extract the \Xir and \Xirb yields.
The model predictions are compared with the \NASixtyOne measurements in Figs.~\ref{fig:ptmodel} and~\ref{fig:dnmodel}. \Epos~1.99  describes well the $\Xir$ and $\Xirb$ transverse momentum and rapidity spectra. The comparison of the \Urqmd~3.4 calculations with the \NASixtyOne measurements reveals significant discrepancies for the $\Xir$ and $\Xirb$ hyperons. The model strongly overestimates $\Xir$ and $\Xirb$ yields. The ratio of $\Xirb$ to $\Xir$ cannot be described by the \Urqmd model but is well reproduced by \Epos~1.99 (see Figs.~\ref{fig:ptmodel} and ~\ref{fig:dnmodel}). 

The statistical model of particle production in various versions has been in use for many years to fit experimental results on particle production 
in \pp as well as 
\textit{N+N} interactions 
(see, e.g. Ref.~\cite{Becattini:2005xt}).
The new measurements by \NASixtyOne of \Xir and \Xirb produced in inelastic \pp interactions at 158~\GeVc as well as previously obtained results for $\pi^+$, $\pi^-$, $K^+$, $K^-$, $p$, $\overline{p}$, $K^*(892)^0$, $\Lambda$, $\phi(1020)$, $\Xi^-$ and $\Xip$ (see Refs.~\cite{Abgrall:2013qoa, Aduszkiewicz:2017sei, Aduszkiewicz:2019zsv, Aduszkiewicz:2020msu, Aduszkiewicz:2019ldi, Aduszkiewicz:2015dmr, Aduszkiewicz:2020dyw}) were compared to  two versions of the Hadron Resonance Gas Model (HRG) using the software package THERMAL-FIST 1.3 of Ref.~\cite{Vovchenko:2019pjl}. For the small \pp system, the appropriate approach is to use the Canonical Ensemble.
The following HRG versions were considered:
\begin{enumerate}[(i)]
\item Canonical Ensemble with fixed strangeness saturation factor, $\gamma_{s} = 1$. The fit parameters are the freeze-out temperature of the particle composition $T$ and 
the fireball radius at freeze-out $R$
\item Canonical Ensemble with the free $\gamma_{s}$ parameter. The fit parameters are $\gamma_{s}$ and $R$. 
\end{enumerate}

\begin{figure}
\centering
\includegraphics[width=.42\textwidth]{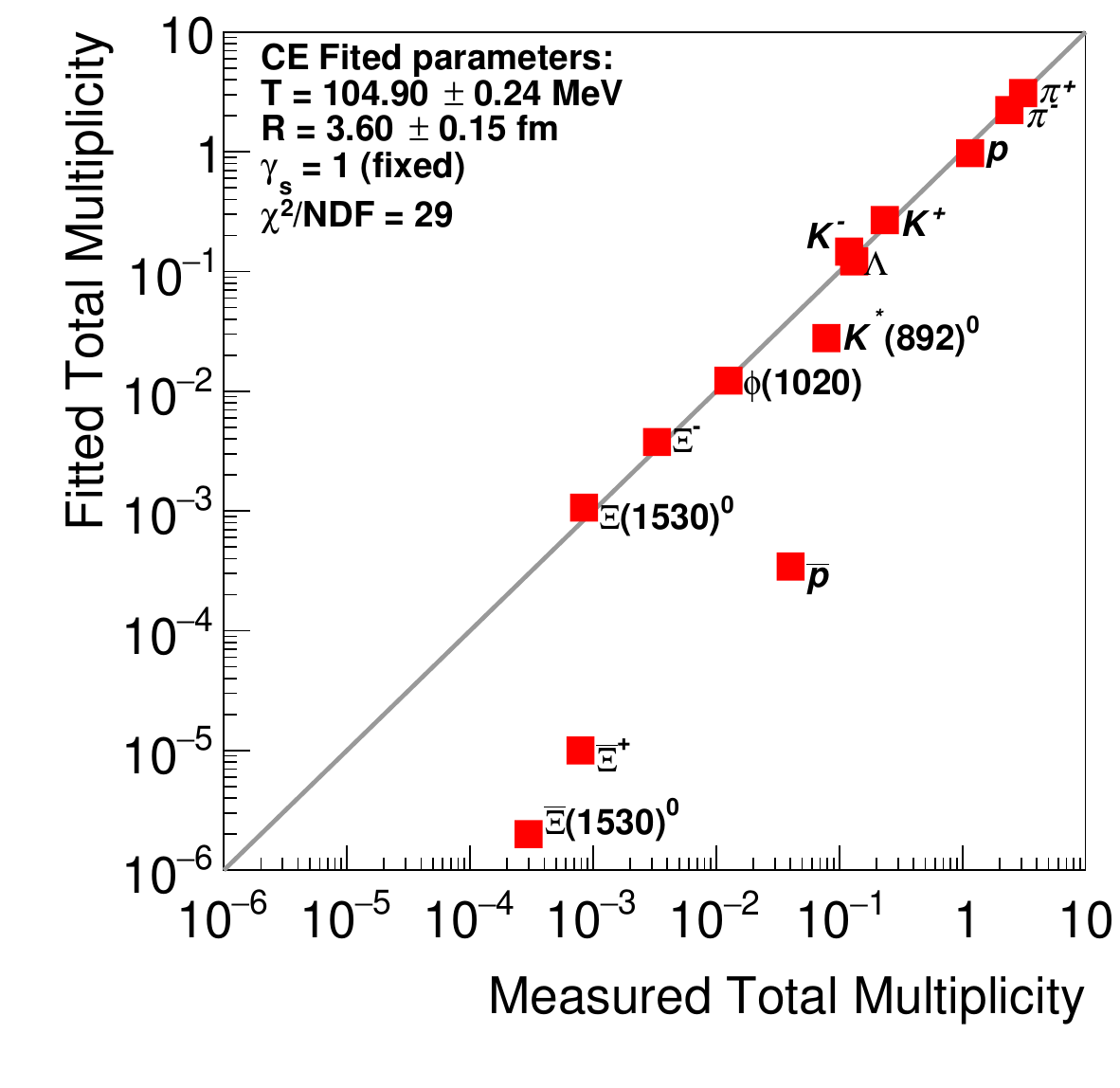}\llap{\parbox[b]{0.05\textwidth}{(\textit{i})\\\rule{0ex}{0.65in}}}
\includegraphics[width=.42\textwidth]{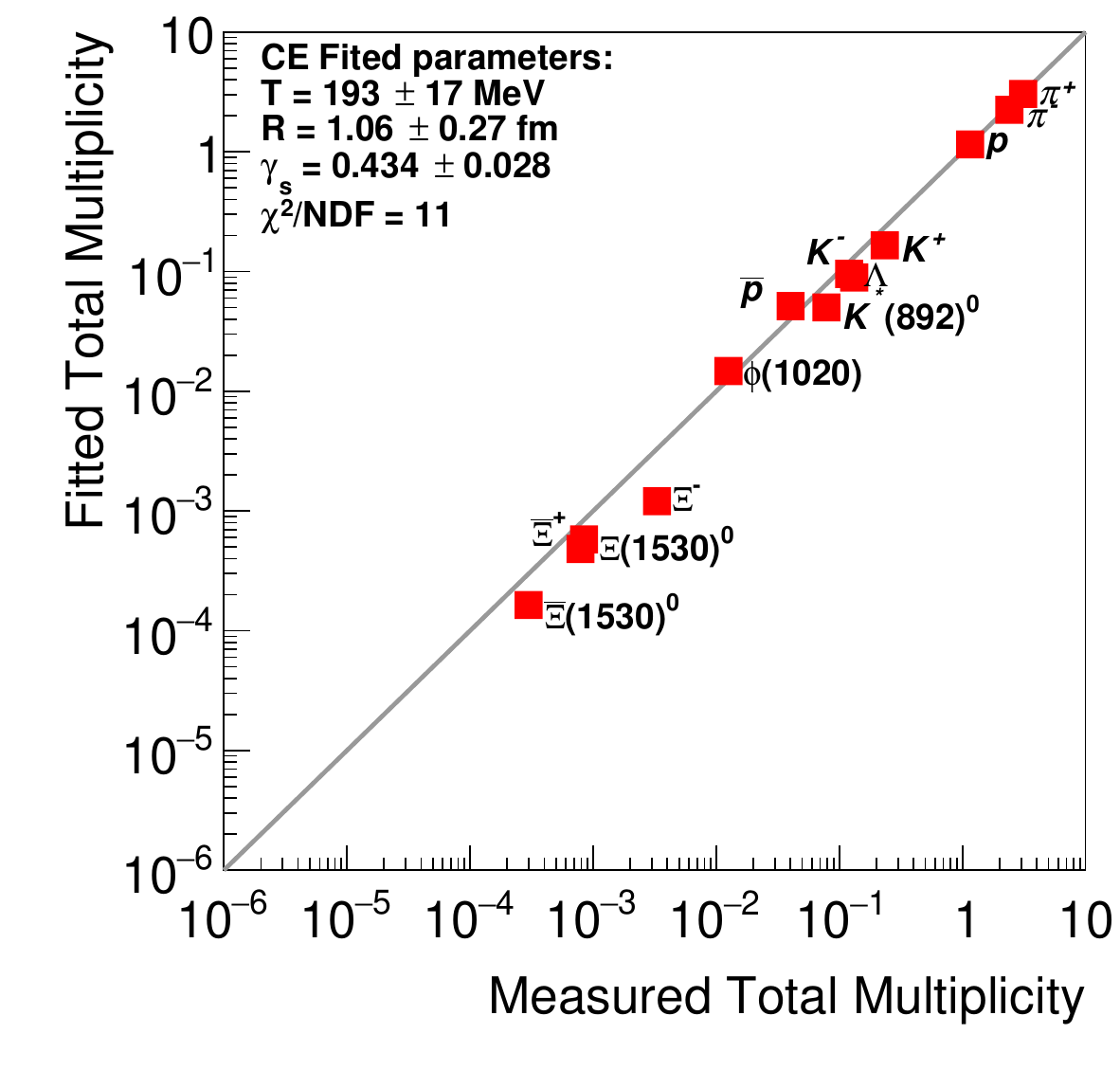}\llap{\parbox[b]{0.05\textwidth}{(\textit{ii})\\\rule{0ex}{0.65in}}}

\caption{(Color online)
Mean multiplicities of $\pi^+$, $\pi^-$, $K^+$, $K^-$, $p$, $\overline{p}$, $K^*(892)^0$, $\Lambda$, $\phi(1020)$, $\Xi^-$, $\Xip$, $\Xir$ and \Xirb produced in \pp interactions at 158~\GeVc~\cite{Abgrall:2013qoa, Aduszkiewicz:2017sei, Aduszkiewicz:2019zsv, Aduszkiewicz:2020msu, Aduszkiewicz:2019ldi, Aduszkiewicz:2015dmr, Aduszkiewicz:2020dyw} measured by \NASixtyOne are compared with mean multiplicities obtained from the HRG model based on the Canonical Ensemble with fixed $\gamma_s = 1$ (\textit{i}) and fitted $\gamma_s$ (\textit{ii}). Uncertainties of the measurements are smaller than the size of the markers.}
\label{fig:HRGmodel}
\end{figure}

Figure~\ref{fig:HRGmodel} compares the measured multiplicities of particles produced in
inelastic \pp interactions at 158~\GeVc with predictions of the two versions of HRG model.
The version with $\gamma_s$ fixed to one shows unacceptably large $\chi^2$/NDF = 29. 
The version with free $\gamma_s$ yields the best fit for $\gamma_s = 0.434 \pm 0.028$.
The description of the data improves, but the $\chi^2$/NDF = 11 is still large.

\section{Summary} \label{sec:summary}

Measurements by \NASixtyOne of double-differential spectra and mean multiplicities of $\Xir$ and $\Xirb$ resonances produced in inelastic \pp interactions were presented. The results were obtained from a sample of 26$\cdot$10$^6$ minimum-bias events at the CERN SPS using a proton beam of 158~\GeVc momentum ($\sqrt{s_{NN}}$ = 17.3~\GeV). 
The measured rapidity and transverse momentum distributions were extrapolated to full phase space, and mean multiplicities of $\Xir $ (6.73 $\pm$ 0.25 $\pm$ 0.67)$\times10^{-4}$ and of $ \Xirb $ (2.71 $\pm$ 0.18 $\pm$ 0.18)$\times10^{-4}$ were obtained. The $\Xirb$/$\Xir$ ratio at mid-rapidity was found to be 0.54 $\pm$ 0.07 $\pm$ 0.08.

The \NASixtyOne results were compared with predictions of two hadronic models \Urqmd and \Epos.
\Epos describes well the experimental results. However, \Urqmd overestimates the yields by a factor of 2.5.
Results were also compared with predictions of the hadron-resonance gas model in the canonical formulation. The equilibrium version of the model is in strong disagreement with the data. This disagreement is slightly reduced by allowing for out-of-equilibrium strangeness production. 

\section*{Acknowledgments}
We would like to thank the CERN EP, BE, HSE and EN Departments for the
strong support of NA61/SHINE.

This work was supported by
the Hungarian Scientific Research Fund (grant NKFIH 123842\slash123959),
the Polish Ministry of Science
and Higher Education (grants 667\slash N-CERN\slash2010\slash0,
NN\,202\,48\,4339 and NN\,202\,23\,1837), the National Science Centre Poland(grants
2014\slash14\slash E\slash ST2\slash00018,
2014\slash15\slash B\slash ST2 \slash\- 02537 and
2015\slash18\slash M\slash ST2\slash00125,
2015\slash 19\slash N\slash ST2 \slash01689,
2016\slash23\slash B\slash ST2\slash00692,
2016\slash21\slash D\slash ST2\slash01983, 
DIR\slash WK\slash\- 2016\slash 2017\slash\- 10-1,
2017\slash\- 25\slash N\slash\- ST2\slash\- 02575,
2018\slash 30\slash A\slash ST2\slash 00226,
2018\slash 31\slash G\slash ST2\slash 03910,
2019\slash 34\slash H\slash ST2\slash 00585),
WUT ID\slash UB,
the Russian Science Foundation, grant 16-12-10176 and 17-72-20045,
the Russian Academy of Science and the
Russian Foundation for Basic Research (grants 08-02-00018, 09-02-00664
and 12-02-91503-CERN),
the Russian Foundation for Basic Research (RFBR) funding within the research project no. 18-02-40086,
the National Research Nuclear University MEPhI in the framework of the Russian Academic Excellence Project (contract No.\ 02.a03.21.0005, 27.08.2013),
the Ministry of Science and Higher Education of the Russian Federation, Project "Fundamental properties of elementary particles and cosmology" No 0723-2020-0041,
the European Union's Horizon 2020 research and innovation programme under grant agreement No. 871072,
the Ministry of Education, Culture, Sports,
Science and Tech\-no\-lo\-gy, Japan, Grant-in-Aid for Sci\-en\-ti\-fic
Research (grants 18071005, 19034011, 19740162, 20740160 and 20039012),
the German Research Foundation DFG (grants GA\,1480/8-1 and project 426579465), the
Bulgarian Nuclear Regulatory Agency and the Joint Institute for
Nuclear Research, Dubna (bilateral contract No. 4799-1-18\slash 20),
Bulgarian National Science Fund (grant DN08/11), Ministry of Education
and Science of the Republic of Serbia (grant OI171002), Swiss
Nationalfonds Foundation (grant 200020\-117913/1), ETH Research Grant
TH-01\,07-3 and the Fermi National Accelerator Laboratory (Fermilab), a U.S. Department of Energy, Office of Science, HEP User Facility managed by Fermi Research Alliance, LLC (FRA), acting under Contract No. DE-AC02-07CH11359 and the IN2P3-CNRS (France).

\bibliographystyle{na61Utphys}
\bibliography{na61References}
\newpage
{\Large The \NASixtyOne Collaboration}
\bigskip
\begin{sloppypar}

\noindent
A.~Acharya$^{\,9}$,
H.~Adhikary$^{\,9}$,
K.K.~Allison$^{\,25}$,
N.~Amin$^{\,5}$,
E.V.~Andronov$^{\,21}$,
T.~Anti\'ci\'c$^{\,3}$,
V.~Babkin$^{\,19}$,
Y.~Balkova$^{\,14}$
M.~Baszczyk$^{\,13}$,
S.~Bhosale$^{\,10}$
% ...

\end{sloppypar}

\noindent
% ...
$^{3}$~Ru{\dj}er Bo\v{s}kovi\'c Institute, Zagreb, Croatia\\
% ...
$^{5}$~Karlsruhe Institute of Technology, Karlsruhe, Germany\\
% ...
$^{9}$~Jan Kochanowski University in Kielce, Poland\\
$^{10}$~Institute of Nuclear Physics, Polish Academy of Sciences, Cracow, Poland\\
% ...
$^{13}$~AGH - University of Science and Technology, Cracow, Poland\\
$^{14}$~University of Silesia, Katowice, Poland\\
% ...
$^{19}$~Joint Institute for Nuclear Research, Dubna, Russia\\
% ...
$^{21}$~St. Petersburg State University, St. Petersburg, Russia\\
% ...
$^{25}$~University of Colorado, Boulder, USA\\
% ...

\end{document}